\documentclass[10pt]{article}

\usepackage{bm}

\usepackage{fullpage}  % for narrow margins

\usepackage{url}
\usepackage[ruled,linesnumbered]{algorithm2e}

\usepackage{multirow} %//multirow for format of table
\usepackage{hhline}  % for \hhline in tables
\usepackage{verbatim}
\usepackage{graphicx}
\usepackage{amsmath,amssymb,amsfonts,graphicx,amsthm,mathtools,nicefrac}%},mathrsfs}
\usepackage{lscape}
\usepackage{color}
\usepackage{authblk}
\usepackage{enumerate}
\usepackage{subcaption}
\usepackage{booktabs}

\usepackage{tikz}
\usetikzlibrary{calc}

\usepackage{pifont}       % \ding{xx}
\usepackage{bbding}       % \Checkmark,\XSolid,... (需要和pifont宏包共同使用)
\usepackage{fontawesome}  % \faCheck,\faTimes

\newcommand{\overbar}[1]{\mkern 1.5mu\overline{\mkern-1.5mu#1\mkern-1.5mu}\mkern 1.5mu}

\newcommand{\KL}{\operatorname{KL}}
\newcommand{\distT}{\operatorname{dist}_{\mathbb T}}
\allowdisplaybreaks
\newcommand\numberthis{\addtocounter{equation}{1}\tag{\theequation}}
\newtheorem{theorem}{Theorem}[section]
\newtheorem{assumption}{Assumption}[section]
\newtheorem{remark}{Remark}[section]

\numberwithin{equation}{section}

\newtheorem{lemma}[theorem]{Lemma}

\DeclareMathOperator*{\argmin}{arg\,min}
\DeclareMathOperator*{\argmax}{arg\,max}

\DeclareMathOperator{\sign}{sign}

\def \lb{\left(}
\def \rb{\right)}
\newcommand{\matsnorm}[2]{\left\| #1\right\|_{{#2}}}

\newcommand{\fronorm}[1]{\ensuremath{\matsnorm{#1}{\footnotesize{\mathsf{F}}}}}
\newcommand{\opnorm}[1]{\ensuremath{\matsnorm{#1}{}}}

\newcommand{\bfm}[1]{\bm{#1}}

\renewcommand{\Pr}[2][]{\mathbb{P}_{#1} \left\{ #2 \rule{0mm}{3mm}\right\}}

\newcommand{\E}[2][]{\mathbb{E}_{#1} \left\{ #2 \rule{0mm}{3mm}\right\}}

\newcommand{\innerR}[2]{\left\langle #1,#2\right\rangle_{\R}}
\def\va{\bfm a}   \def\mA{\bfm A}  
\def\vb{\bfm b}   \def\mB{\bfm B}  
   \def\mC{\bfm C}  \def\C{\mathbb{C}}

\def\vg{\bfm g}     
\def\vh{\bfm h}   \def\mH{\bfm H}  
   \def\mI{\bfm I}  
   \def\mJ{\bfm J}

   \def\mM{\bfm M}

   \def\mQ{\bfm Q}  
     \def\R{\mathbb{R}}
\def\vs{\bfm s}   \def\mS{\bfm S}  
     
\def\vu{\bfm u}   \def\mU{\bfm U}  
\def\vv{\bfm v}   \def\mV{\bfm V}  
\def\vw{\bfm w}     
\def\vx{\bfm x}   \def\mX{\bfm X}  
\def\vy{\bfm y}   \def\mY{\bfm Y}  
\def\vz{\bfm z}   \def\mZ{\bfm Z}

\def \tranH{\mathsf{H}}

\usepackage{bbm}

\def \hh{\widehat{\vh}}

\makeatletter
\@removefromreset{assumption}{section}   % 取消每当 section 增加时重置 assumption 计数器
\makeatother

\makeatletter
\ifx\c@theorem\undefined\else\expandafter\let\csname c@theorem\endcsname\relax\fi

\ifx\c@lemma\undefined\else\expandafter\let\csname c@lemma\endcsname\relax\fi

\newtheorem{theorem}{Theorem}[section]
\newtheorem{lemma}{Lemma}[section]
\makeatother

\begin{document}

\title{Non-Asymptotic Performance Guarantees for MLE under a Gumbel–Softmax Index Model with Phaseless Measurements}
\author[1]{Jinchi Chen}
\author[2]{Mingxi Hu}
\author[3]{Peigang Jiang}
\author[3]{Xin Meng}
\author[2]{Ke Wei}
\author[2]{Xianyin Zhang}
\affil[1]{School of Mathematics, East China University of Science and Technology, Shanghai,
China.}
\affil[2]{School of Data Science, Fudan University, Shanghai, China}
\affil[3]{Huawei Technologies Co., Ltd., Shanghai, China}
\date{\today}

%%%%%%
\maketitle
%%%%%%
\begin{abstract}
This paper studies the recovery of a complex-valued signal from index-only
observations generated by a phaseless Gumbel--Softmax model. The model is
motivated by limited-feedback frequency-division duplex massive MIMO
systems, where it provides a smooth probabilistic surrogate for the
deterministic hard-PMI selection rule. We consider a constrained maximum
likelihood estimator and establish 
non-asymptotic statistical guarantees for its recovery performance. Under a bounded-design condition, we first derive a global excess-risk bound
of order $\sqrt{d/T}$, which does not explicitly depend on the number of
candidate indices. A quantitative local identifiability condition is then
introduced to characterize the regime in which sharper local guarantees are
available. Under this condition, the excess risk and squared parameter error
both scale as $d/T$. The condition is further verified with high
probability for a Haar--Stiefel random design, and a local minimax lower bound
is established that matches the upper bound in its dependence on the sample
size, signal dimension, temperature parameter, and signal norm, up to
design-dependent factors. Numerical experiments on  FDD 
downlink channel estimation demonstrate that the proposed likelihood-based
approach can overall achieve superior reconstruction performance, supporting the
effectiveness of the Gumbel–Softmax model as a surrogate for hard PMI
feedback.
\end{abstract}
%%%%%%%%%%%%%%%%%
\section{Introduction}
%\subsection{Problem formulation}
%\label{sec:problem-formulation}
In this paper, we study the problem of recovering a signal, up to a global phase, from index-valued observations generated by the \emph{phaseless Gumbel--Softmax index model}. Here the goal is to recover an unknown signal \(\vh\in\C^d\setminus\{\bm 0\}\) from $T$ independent random indices
 \begin{align}
	I_t
	&=
	\operatorname*{arg\,max}_{i\in[N]}
	\left\{
	\frac{|\va_{t,i}^{\tranH}\vh|^2}{\tau}
	+g_{t,i}
	\right\},\quad t=1,\cdots,T,
	\label{eq:intro-observation-model}
\end{align}
where \(\tau>0\) is a fixed temperature parameter,
\(\{g_{t,i}\}_{t\in[T],\,i\in[N]}\) are independent and identically
distributed \(\mathrm{Gumbel}(0,1)\) random variables with density
\[
f(x)=\exp\!\left[-\left(x+e^{-x}\right)\right],
\quad x\in\R,
\]
and \(\{\va_{t,i}, i=1,\cdots,N\}_{t=1}^T\) are measurement vectors. By the Gumbel-max trick, the random
index \(I_t\) follows a categorical distribution \cite{Gumbel1954}. That is, for every \(i\in[N]\) and every candidate signal
\(\vx\in\C^d\), if we define 
\begin{align*}
	p_t(i;\vx)
	:=
	\frac{\exp\!\left(|\va_{t,i}^{\tranH}\vx|^2/\tau\right)}
	{\sum_{j=1}^N\exp\!\left(|\va_{t,j}^{\tranH}\vx|^2/\tau\right)},
\end{align*}
then
\(
	\Pr{I_t=i}
	=p_t(i;\vh).
\)

The primary motivation for studying the observation model in \eqref{eq:intro-observation-model} is that it serves as a smooth probabilistic surrogate for the deterministic hard-PMI selection rule in a frequency-division duplex (FDD) massive MIMO system. Consider the single-antenna user setting, where the base station has $d$ antennas and the user has a single antenna. Thus the downlink channel can be represented by a vector $\vh \in \C^d$. In the limited-feedback FDD system, the base station and the user interact over multiple communication rounds, each consisting of a measurement phase and a feedback phase. More precisely,
at the $t$-th communication round, the base station uses a prescribed dimension-reduction matrix $\mQ_t \in \C^{d\times p}$ and transmits pilots, from which the user estimates the effective \(\mQ_t^\mathsf{H}\vh\) . Assume that the base station and the user share a codebook
\begin{align*}
    \mV =
    \begin{bmatrix}
        \vv_1 & \cdots & \vv_N
    \end{bmatrix}
    \in \C^{p\times N},\numberthis\label{eq:codewords}
\end{align*}
whose columns are candidate precoding vectors in the reduced-dimensional domain. The user then feeds back a precoding matrix indicator (PMI), which is the index of the codeword that maximizes the effective channel gain:
\begin{equation}
    i_t = \argmax_{i=1,\dots,N} \big| \vv_i^\tranH \mQ_t^\tranH \vh \big|^2,\quad t=1,\cdots, T.
    \label{eq:pmi}
\end{equation}
An idealized unquantized channel quality indicator (CQI) associated with the selected codeword is given by
\begin{equation}
    \eta_t = \big| \vv_{i_t}^\tranH \mQ_t^\tranH \vh \big|^2,\quad t=1,\cdots, T.\label{eq:CQI}
\end{equation}
Depending on the feedback configuration, the base station may estimate the channel from either the PMI sequence alone or the joint PMI/CQI feedback. Define 
\begin{align*}
 \va_{t,i}:=\mQ_t\vv_i\in\C^d.\numberthis\label{eq:FDDa}
\end{align*}
For the more challenging setting where the base station only observes the PMI sequence $\{i_t\}_{t=1}^T$, \eqref{eq:intro-observation-model} provides a smooth surrogate model since the softmax distribution tends to a one‑hot vector associated with the PMI feedback as $\tau\to0^+$. This probabilistic smoothing yields a differentiable likelihood and leads naturally to a constrained maximum likelihood estimator that also works well even when the indices are obtained from the hard PMI model, as demonstrated by the numerical results.

\subsection{Related Work}

% 讲相位恢复的 (稀疏相位恢复的一笔带过)
\paragraph{Phase retrieval}
The recovery problem studied in this paper is closely related to phase
retrieval, namely the problem of recovering a signal $\vh\in\C^d$ from
magnitude-only measurements of the form $\{|\va_i^{\tranH}\vh|^2\}_{i=1}^m$. Phase retrieval originates from optical imaging, where the classical
alternating-projection algorithms of Gerchberg--Saxton and Fienup remain
widely used \cite{gerchberg1972practical,fienup1982phase}; see
\cite{shechtman2015phase,fannjiang2020numerics} for contemporary overviews of
its applications and numerics. A modern line of work establishes recovery
guarantees under random measurement models. PhaseLift
\cite{candes2013phaselift} lifts the signal to the rank-one matrix
$\vh\vh^{\tranH}$ and solves a semidefinite relaxation, achieving exact and
stable recovery from $O(d\log d)$ Gaussian measurements. Subsequent nonconvex
methods work directly in the signal space, including alternating minimization
\cite{altmin_netrapalli2013phase}, Wirtinger flow and its truncated variant
\cite{candes2015phase,TWF_chen2015}, and truncated amplitude flow
\cite{wang2018taf}. These methods typically combine a spectral initialization with iterative
refinement and provably converge at a linear rate from $O(d\log d)$, or even
$O(d)$, measurements. Beyond carefully initialized schemes,
\cite{sun2018geometric} shows that the natural least-squares objective has a
benign global geometry once the sample size exceeds the order of
$d\log^3 d$, and \cite{chen2019gradient} proves that plain gradient descent
with random initialization succeeds with near-optimal sample and iteration
complexity. When the underlying signal is sparse, the sample complexity can
be further reduced by exploiting the sparsity prior, and a parallel line of
work develops convex and nonconvex sparse phase retrieval algorithms with
provable guarantees, see for example
\cite{ohlsson2012cprl,cai2016optimal,wang2017sparse,jagatap2019sample,cai2022sparse,cai2024a,xu2024subspace};

There is also a line of research on one-bit phase retrieval, where each observation is the sign of the
difference between a designated pair of squared-magnitude measurements,
$\sign\lb|\va_{i,1}^{\tranH}\vx|^2-|\va_{i,2}^{\tranH}\vx|^2\rb$, and
recovery is carried out by greedy and alternating-minimization refinements
that are provably robust to unknown nonlinear distortions of the
measurements \cite{mroueh2013quantization}. The work \cite{mukherjee2018binary} instead quantizes each
quadratic measurement against a fixed positive threshold and develops a
lifting-based projected-gradient method that enforces consistency with the
binary data, while \cite{eamaz2022onebit} studies phase retrieval from
one-bit samples acquired by time-varying threshold comparators, trading
quantization resolution for very high sampling rates. Recently,
\cite{chen2026onebit} established minimax-optimal error rates for one-bit
phase retrieval from Gaussian measurements quantized as
$\sign(|\va_i^{\tranH}\vx|-\tau_i)$, for both unstructured and sparse
signals, together with efficient spectrally initialized algorithms attaining
these rates. In all of these settings, each individual measurement, or each
designated pair of measurements, is quantized separately according to a rule
specified by the designer.  In contrast to
one-bit phase retrieval, each round of the sample model considered in this paper only reveals a single index
$I_t$---at most $\log_2N$ bits---produced by a competition among the $N$
codewords fixed by the protocol, rather than by per-measurement thresholds
chosen by the estimator.

\paragraph{Estimation from categorical observations}
Beyond phase retrieval, our observation model is also connected to
statistical estimation from categorical observations of a structured
parameter. In categorical, or multinomial, matrix completion
\cite{cao2015categorical,klopp2015adaptive}, one observes categorical
responses generated from a low-rank matrix through a link function and
recovers the matrix by constrained likelihood maximization, with
non-asymptotic error bounds; after the rank-one lifting
$\vh\mapsto\vh\vh^{\tranH}$, our model has precisely this structure, since
the conditional distribution of $I_t$ depends on the linear functionals
$\innerR{\mA_{t,i}}{\vh\vh^{\tranH}}$ through a softmax link. From another
angle, by the Gumbel-max equivalence, the observation
\eqref{eq:intro-observation-model} is a multinomial-logit
(Plackett--Luce-type) choice among $N$ alternatives with utilities
$|\va_{t,i}^{\tranH}\vh|^2/\tau$, which connects our model to
maximum-likelihood estimation from pairwise-comparison and partial-ranking
data, for which sharp non-asymptotic and minimax theory is available
\cite{negahban2017rank,hajek2014minimax}. The same softmax smoothing also
underlies the Gumbel-softmax relaxation widely used in machine learning
\cite{jang2017categorical,maddison2017concrete}.  In contrast to categorical matrix completion, where each observation depends
on the unknown matrix only through a single randomly sampled entry---the
category probabilities being fixed link functions of that entry
\cite{cao2015categorical} or governed by separate per-category parameter
matrices \cite{klopp2015adaptive}---the measurement matrices $\mA_{t,i}$
in our model are deterministic rank-one matrices and the
dimension-reduction matrices, and the $N$ competing logits are distinct
linear functionals of the same lifted matrix $\vh\vh^{\tranH}$. In contrast to the ranking literature, the
utilities are not free parameters but a common phaseless quadratic function
of $\vh$, which is identifiable only up to a global phase. Establishing
non-asymptotic guarantees for the constrained MLE \eqref{eq:MLE} therefore
requires combining the analysis of categorical likelihoods with the quotient
geometry of phase retrieval, which is the subject of this paper.

% FDD基于相位恢复的
\paragraph{Phase-retrieval-based limited feedback in FDD systems}
In the limited-feedback FDD setting described above, a recent line of work
exploits the fact that standardized feedback quantities---the PMI and the
CQI---are phaseless, coarsely quantized functionals of the downlink
channel \cite{3gpp38214v1620_codebook_pmi,ning2026precoding}, and formulates their
utilization at the base station as a phase retrieval problem. In particular,
\cite{Li2023CPR} casts the joint PMI/CQI feedback as a
constrained phase retrieval problem, in which the PMI feedback delineates
the feasible region, and solves it by a two-stage first-order algorithm;
\cite{Li2024CCM} reconstructs the downlink channel covariance matrix from
limited feedback; \cite{Li2024CQI} refines multi-stream beamforming from CQI
feedback through a tuning-free scheme; \cite{zhang2025spectral} models the
downlink channel as a spectrally sparse signal and recovers it from CQI
feedback by nonconvex gradient methods that exploit the induced low-rank
Hankel structure; and \cite{Li2025AEE} develops an
online alternating exploration--estimation procedure that adaptively designs
probing beams and updates the channel estimate from the induced feedback.
These works focus on algorithm design and demonstrate
strong empirical performance, and statistical guarantees relating the number of
feedback rounds to the achievable recovery accuracy are beyond their scope. In contrast,
we addresses the  restrictive form
of feedback, in which only the winning index is reported. Modeling the
feedback by the phaseless Gumbel--Softmax index model
\eqref{eq:intro-observation-model}, we establish non-asymptotic excess-risk
and parameter-error bounds for the corresponding constrained MLE \eqref{eq:MLE}, together
with a matching local minimax lower bound.

% 贡献与文章结构（说清楚针对单天线的情况）
\subsection{Main Contributions}
The goal of this paper is to develop a sharp non-asymptotic theory for the constrained MLE
\eqref{eq:MLE} under the phaseless Gumbel--Softmax index model and to evaluate
the resulting likelihood-based method for PMI-only channel estimation. Our
main contributions are summarized as follows.
\begin{itemize}
    \item \textbf{Non-asymptotic guarantees for the constrained MLE under
    the phaseless Gumbel--Softmax index model.}
     Under a bounded-design condition, the constrained
    MLE achieves a global excess-risk bound of order $\sqrt{d/T}$ for an
    arbitrary bounded collection of measurement vectors
    (Theorem~\ref{thm:global}). Under a  local identifiability condition (Assumption~\ref{assump:fisher-frame}), we further obtain the parametric rate $d/T$ for both the excess risk
    and the squared estimation error modulo the unavoidable global phase
    (Theorem~\ref{thm:local}). The analysis uses a global risk-to-distance
    inequality (Theorem~\ref{thm:risk-to-distance}) to localize the estimator
    before invoking local curvature. We further provide a verifiable
    high-probability certificate for the identifiability condition under a
    Haar--Stiefel random design
    (Lemma~\ref{lem:complex-secant-stiefel}).

    \item \textbf{A matching local minimax lower bound.}
    We derive a local minimax lower bound for estimation under the same
    observation model, with the estimation error measured modulo the global
    phase (Theorem~\ref{thm:minimax-lower}).  Up to design-dependent factors, the MLE upper bound and the minimax lower 
bound exhibit the same dependence on $T$, $d$, $\tau$, and the signal norm, 
thereby establishing the optimality of the local rate.

    \item \textbf{Empirical evaluation with deterministic hard-PMI feedback.}
    We apply the likelihood-based estimator directly to deterministic
    hard-PMI observations obtained from QuaDRiGa-generated FDD channels.
    The experiments 
    consider fully random as well as structured dimension-reduction designs.
    The results show that the proposed method consistently delivers stronger
reconstruction performance than the competing baselines, with beam precision
improving steadily as the number of feedback rounds increases. These results support the practical
    relevance of the Gumbel--Softmax likelihood as a smooth surrogate for
    hard-PMI feedback.
\end{itemize}
%%说一下technical的事情， 如果有regularity，shengyang的邮件

\paragraph{Analytical challenges.}
Under standard local regularity conditions, classical asymptotic likelihood
theory suggests a $1/T$ excess-risk scaling for the MLE. This asymptotic
prediction, however, does not by itself imply a non-asymptotic guarantee of
the same order for the constrained MLE considered here. Although the
observations are independent categorical draws from a smooth parametric
model, independence across communication rounds facilitates concentration
but does not alone ensure the quantitative identifiability and local
curvature required for a finite-sample analysis. In particular, global-phase
invariance makes the Fisher information singular along the phase direction,
so curvature must instead be established on the corresponding horizontal
space. Moreover, because the empirical objective is nonconvex, local
curvature can be exploited only after the constrained MLE has been shown to
lie near the phase orbit of the true signal. A key step in our analysis is to identify a quantitative local identifiability condition, formulated as a two-sided weighted Fisher-frame condition, that provides the structural basis for localization and local curvature. Under this condition, we convert the global excess-risk bound into a risk-to-distance localization guarantee and establish uniform local empirical curvature on the horizontal space. We also show that this condition holds with high probability under Haar--Stiefel probing.

\subsection{Notation and Organization}

\paragraph{Notation}
For a positive integer $n$, let $[n]:=\{1,\ldots,n\}$. We write $\mathrm i$
for the imaginary unit and $\operatorname{Re}(\cdot)$,
$\operatorname{Im}(\cdot)$ for the real and imaginary parts of a complex
quantity. Boldface lowercase letters (e.g., $\vx,\vh$) denote vectors,
boldface uppercase letters (e.g., $\mQ,\mA$) denote matrices,
$(\cdot)^{\tranH}$ denotes the conjugate transpose, and $\mI_p$ is the
$p\times p$ identity matrix. For a vector $\vx$, $\|\vx\|_2$ and
$\|\vx\|_\infty$ denote the Euclidean and entrywise supremum norms; for a
matrix $\mX$, $\fronorm{\mX}$ and $\opnorm{\mX}$ denote the Frobenius and
operator norms. We write $\E{\cdot}$ and $\Pr{\cdot}$ for expectation and
probability; a subscript, as in $\E[\vh]{\cdot}$, indicates the source of
randomness over which the expectation is taken. For probability vectors $p$ and $q$ on
$[N]$, $\KL(p\,\|\,q):=\sum_{i=1}^Np_i\log(p_i/q_i)$ is the
Kullback--Leibler divergence. Throughout, $C,c,C_1,C_2,\ldots$ denote
positive absolute constants whose values may change from line to line, and
$a\lesssim b$ means $a\leq Cb$ for such a constant, and $a\gtrsim b$ is
defined analogously.

\paragraph{Real Hilbert structure and derivatives}
For $\vx,\vy\in\C^d$, and more generally for complex matrices $\mX,\mY$ of
the same size, define
\begin{align}
	\innerR{\vx}{\vy}
	&:=
	\operatorname{Re}\operatorname{tr}(\vx^{\tranH}\vy)
	=
	\operatorname{Re}(\vx^{\tranH}\vy),
	\quad
	\innerR{\mX}{\mY}
	:=
	\operatorname{Re}\operatorname{tr}(\mX^{\tranH}\mY).
	\label{eq:real-inner-product}
\end{align}
The blackboard-bold subscript $\R$ labels the real inner product.
  That is, these definitions amount to identifying a complex
vector (or matrix) with the long real vector obtained by stacking its real
and imaginary parts, e.g., identifying \(\C^d\) with \(\R^{2d}\), under
which \(\innerR{\cdot}{\cdot}\) is the standard Euclidean inner product and
every derivative below is an ordinary real derivative.  The induced norm is still the Euclidean norm for vectors and the Frobenius norm for
matrices. Thus, for a
real-valued function $F$ and a complex direction $\mV$,
\begin{align}
	DF(\mX)[\mV]
	&:=
	\left.\frac{d}{ds}F(\mX+s\mV)\right|_{s=0}
	=
	\innerR{\nabla F(\mX)}{\mV},
	\label{eq:real-gradient-definition}\\
	D^2F(\mX)[\mU,\mV]
	&:=\left.\frac{\partial^2}{\partial s\partial t} F(\mX+s\mU+t\mV)\right|_{s=t=0}=
	\innerR{\nabla^2F(\mX)[\mU]}{\mV}.
	\label{eq:real-hessian-definition}
\end{align}
Here $\nabla^2F(\mX)$ is a real-linear self-adjoint operator.  In particular,
if $\mA$ is Hermitian, then
\begin{align}
	D(\vx^{\tranH}\mA\vx)[\vv]
	=
	2\innerR{\mA\vx}{\vv},
	\quad
	\nabla(\vx^{\tranH}\mA\vx)=2\mA\vx.
	\label{eq:quadratic-real-gradient}
\end{align}

\paragraph{Organization} 
The remainder of this paper is organized as follows. Section~\ref{sec:mle} presents the MLE in a formal way, with an alternative interpretation.
Section~\ref{sec:main-results} contains the main results: a global
excess-risk bound at the slow rate $1/\sqrt T$, the quotient geometry and
 localization analysis, the sharp
local excess-risk bound at the rate $d/T$ together with the corresponding
parameter-error result, and a matching local minimax lower bound.
Section~\ref{sec:numerics} reports the numerical experiments.
Sections~\ref{sec:proofs-main} and~\ref{sec:proofs-lemmas} collect the
proofs of the main theorems and of the technical lemmas, respectively.
 Section~\ref{sec conclusion} concludes
the paper with directions for future work.
%%%%%%%%%%%%%%%%%%%%%%%%
%%%%%%%%%%%%%%%%%%%%%%%%

\section{Maximum Likelihood Estimation (MLE)}\label{sec:mle}
Define
\begin{align*}
	\ell_t(i;\vx)
	=
	-\log p_t(i;\vx)
	=
	\log\sum_{j=1}^N
	\exp\left(
	\frac{|\va_{t,j}^\tranH \vx|^2-|\va_{t,i}^\tranH \vx|^2}{\tau}
	\right).\numberthis\label{eq:defoflt}
\end{align*}
Then, given $T$ independent observations $I_1,\cdots,I_T$ from the Gumbel-Softmax model \eqref{eq:intro-observation-model}, the negative log-likelihood function is given by
\begin{align}
\label{negative log-likelihood}
L_T(\vx):=\frac{1}{T}\sum_{t=1}^T \ell_t(I_t;\vx) = \frac{1}{T}\sum_{t=1}^T\log\sum_{j=1}^N
	\exp\left(
	\frac{|\va_{t,j}^\tranH \vx|^2-|\va_{t,i}^\tranH \vx|^2}{\tau}
	\right).
\end{align}
%Let $\calT:=\{\vx\in\R^d:\|\vx\|_2\le R\}$. 
We consider the following constrained maximum likelihood approach for the estimation of $\vh$,
\begin{align*}
	\hh_T\in\argmin_{\vx: \|\vx\|_2\leq R} L_T(\vx),\numberthis\label{eq:MLE}
\end{align*}
where $R>0$ is a fixed radius parameter. The constraint $\|\vx\|_2\leq R$ is introduced to restrict the parameter space to a compact set. This is without loss of generality, since the channel vector $\vh$ has finite energy in practical systems, and $R$ can be chosen sufficiently large such that the true channel lies in this set. 

Beyond its probabilistic derivation as a negative log-likelihood,
the objective $L_T$ also admits a complementary interpretation as
a smooth surrogate for empirical decision-error minimization under
the deterministic hard-PMI rule. To make this connection precise,
consider first the empirical $0$--$1$ loss associated with the
hard-PMI rule,
\begin{align*}
	L_T^{0-1}(\vx) :=	\frac{1}{T}\sum_{t=1}^T	1\Big\{
	I_t \notin \arg\max_{j\in[N]} |\va_{t,j}^\tranH \vx|^2
	\Big\},
\end{align*}
Noting that
\begin{align*}
	I_t \notin \arg\max_{j\in[N]} |\va_{t,j}^\tranH \vx|^2 \Longleftrightarrow  \max_{j\in[N]}|\va_{t,j}^\tranH \vx|^2 -|\va_{t,I_t}^\tranH \vx|^2>0.
\end{align*}
Therefore, $L_T^{0-1}(\vx)$ can be equivalently rewritten as
\begin{align*}
	L_T^{0-1}(\vx) =  \frac{1}{T}\sum_{t=1}^T
	1\Big\{	\max_{j\in[N]}|\va_{t,j}^\tranH \vx|^2 -|\va_{t,I_t}^\tranH \vx|^2>0.
	\Big\}.
\end{align*}
Though it is the most direct formulation of empirical error minimization under the hard PMI rule, the objective is discontinuous and difficult to optimize.

We now smooth this objective in two steps. First, replacing the discontinuous indicator loss by the margin yields
\begin{align*}
L^{\mathrm{margin}}(\vx)=\frac{1}{T}\sum_{t=1}^T\left(\max_{j\in[N]}|\va_{t,j}^\tranH \vx|^2 -|\va_{t,I_t}^\tranH \vx|^2\right),\quad\vx\neq 0.
\end{align*}
This loss function is continuous, but  still contains the nonsmooth max operator.
Thus, we further replace the $\mathrm{max}$ operator by its log-sum-exp
smooth approximation,
\[
\max\{x_1,\cdots,x_N\}\rightarrow \tau\log\left(\frac{1}{N}\sum_{i=1}^N\exp(x_i/\tau)\right),
\]
leading to
\[
L_T(\vx)=\frac{1}{T}\sum_{t=1}^T\left(\tau\log \sum_{j=1}^N \exp(|\va_{t,j}^\mathsf{H}\vx|^2/\tau)-|\va_{t,I_t}^\mathsf{H}\vx|^2\right)-\tau\log N,
\]
which  coincides with the negative log-likelihood up to a positive scaling factor and an additive constant.

%%%%%%%%%%%%%%%%%%%%%%%%
%%%%%%%%%%%%%%%%%%%%%%%%
\section{Non-asymptotic performance guarantee of MLE}
\label{sec:main-results}
% excessive risk定义
% 接下来给出慢的1/sqrt{T}
% 交叉着来
% 最后give minimax 
Let $\overbar L_T(\vx):=\E{L_T(\vx)}$ and define the excess risk as 
\begin{align}\label{eq:excess risk}
\overbar{L}_T(\vx)-\overbar L_T(\vh).
\end{align}
A direct calculation yields that 
\begin{equation}
	\label{eq:pop-kl}
	\overbar L_T(\vx)-\overbar L_T(\vh)
	=
	\frac{1}{T}\sum_{t=1}^T
	\mathrm{KL}\bigl(p_t(\cdot;\vh)\,\|\,p_t(\cdot;\vx)\bigr),
\end{equation}
where $\mathrm{KL}\bigl(\cdot\,\|\,\cdot\bigr)$ denotes the KL divergence of two probability distributions.
Next, we will establish the global and local excess risks for the MLE, as well as a local minimax lower bound for the estimation of $\vh$.
\subsection{Global excess-risk bound}

The global excess-risk bound follows from a uniform law of large numbers over the complex
parameter ball.  The argument uses only boundedness, independence, and the
quadratic-softmax structure.

\begin{assumption}[Bounded complex design]
\label{assump:bounded-design}
The true channel satisfies $0<\|\vh\|_2\leq R$, and
\begin{align*}
	\|\va_{t,i}\|_2\leq1,
	\quad t\in[T],\quad i\in[N].
\end{align*}
The latter condition follows, for example, when $\|\vv_i\|_2=1$ and
$\mQ_t^{\tranH}\mQ_t=\mI_p$ when $\va_{t,i}$ is given by \eqref{eq:FDDa} in the FDD problem.
\end{assumption}

\begin{theorem}[Global excess-risk bound]
\label{thm:global}
Under Assumption~\ref{assump:bounded-design}, there are absolute constants
$C_1,C_2>0$ such that, for every $\gamma\in(0,1)$, with probability at least
$1-\gamma$,
\begin{align*}
	\overbar L_T(\widehat{\vh}_T)-\overbar L_T(\vh)
	\leq
	\frac{C_1R^2\sqrt d}{\tau\sqrt T}
	+
	\frac{C_2R^2}{\tau\sqrt T}\sqrt{\log(1/\gamma)}.
\end{align*}
\end{theorem}

\subsection{Quotient geometry and localization}
It is evident that the likelihood depends on a channel vector only through its rank-one lift. Thus,
the vectors $\vx$ and $e^{-\mathrm i\phi}\vx$ are observationally equivalent
for every rotation angle $\phi\in[0,2\pi)$.  Define the quotient distance
\begin{align*}
	\distT(\vx,\vh)
	:=
	\min_{\phi\in[0,2\pi)}
	\|e^{-\mathrm i\phi}\vx-\vh\|_2.
\end{align*}
Expanding the square and optimizing over $\phi$ gives the closed form
\begin{align}
	\distT^2(\vx,\vh)
	=
	\|\vx\|_2^2+\|\vh\|_2^2-2\,|\vh^{\tranH}\vx|.
	\label{eq:distT-closed}
\end{align}
We align each phase orbit with $\vh$ and remove the infinitesimal direction that changes only the global phase. More precisely, the phase orbit through $\vh$ has real tangent space
$\operatorname{span}_{\R}\{\mathrm i\vh\}$.  We therefore restrict local
perturbations to its real-orthogonal complement,
\begin{align*}
	\mathcal H_{\vh}
	:=
	\left\{
	\vz\in\C^d:
	\innerR{\mathrm i\vh}{\vz}=0
	\right\},
\end{align*}
which is known as the horizontal space at $\vh$.  Since
$\innerR{\mathrm i\vh}{\vz}
=\operatorname{Im}(\vh^{\tranH}\vz)$, this condition is equivalent to
$\vh^{\tranH}\vz\in\R$.  %Lemma~\ref{lem:phase-procrustes-alignment} below shows that an optimally
%phase-aligned representative has precisely such a horizontal displacement
%from $\vh$.

In addition, define the rank-one measurement matrices
\begin{align}
	\label{eq A}
	\mA_{t,i}:=\va_{t,i}\va_{t,i}^{\tranH},
	\quad t\in[T],\quad i\in[N].
\end{align}
Then there holds \(|\va_{t,i}^{\tranH}\vx|^2=\innerR{\mA_{t,i}}{\vx\vx^{\tranH}}\)
for every \(\vx\in\C^d\). For a Hermitian matrix $\mM$, define
\begin{align*}
	q_t(i;\mM)
	:=
	\frac{\exp\!\left(\innerR{\mA_{t,i}}{\mM}/\tau\right)}
	{\sum_{j=1}^N\exp\!\left(\innerR{\mA_{t,j}}{\mM}/\tau\right)}.
\end{align*}

	\begin{assumption}[Two-sided weighted Fisher frame]
		\label{assump:fisher-frame}
		There are constants $0<\kappa_F^-\leq\kappa_F^+<\infty$ such that, for every Hermitian matrix $\mM$ admitting a representation
		\begin{align*}
			\mM =		\sum_{k=1}^{K}\lambda_k\vx_k\vx_k^{\tranH},~\lambda_k\geq0, ~\sum_{k=1}^{K}\lambda_k=1,~ \|\vx_k\|_2\leq R,
		\end{align*}
		and every
		$\bm\Delta=\vx\vx^{\tranH}-\vy\vy^{\tranH}$
		with $\|\vx\|_2,\|\vy\|_2\leq R$, 
		\begin{align}
			\kappa_F^-\fronorm{\bm\Delta}^2
			&\leq
			\frac1T\sum_{t=1}^T\sum_{i,j=1}^N
			q_t(i;\mM)q_t(j;\mM)
			\innerR{\mA_{t,i} - \mA_{t,j}}{\bm\Delta}^2
			\leq
			\kappa_F^+\fronorm{\bm\Delta}^2.
			\label{eq:fisher-frame}
		\end{align}
		%The lower constant is used for the	MLE upper bound, whereas the upper constant controls KL divergence in the	minimax lower bound.
	\end{assumption}

	When $\va_{t,i}$ is given by \eqref{eq:FDDa} in the FDD problem, the following lemma gives a concrete random-design model under which that condition holds with high probability. For $p\leq d$, let
	\begin{align*}
		\operatorname{St}(d,p)
		:=
		\{\mQ\in\C^{d\times p}:\mQ^{\tranH}\mQ=\mI_p\}
	\end{align*}
	be the complex Stiefel manifold with Haar probability measure, and define
	the projective codebook spread
	\begin{align}
		s_{\mathcal V}
		:=
		\frac1{N^2}\sum_{i,j=1}^N
		\fronorm{\vv_i\vv_i^{\tranH}-\vv_j\vv_j^{\tranH}}^2.
		\label{eq:codebook-spread}
	\end{align}
	
	\begin{lemma}[Complex Haar--Stiefel weighted Fisher frame]
		\label{lem:complex-secant-stiefel}
		Suppose that $d\geq2$, $N\geq2$, $p\leq d$, and that the codewords
		$\vv_1,\ldots,\vv_N\in\C^p$ in \eqref{eq:codewords} satisfy
		\begin{align*}
			\|\vv_i\|_2=1,
			\quad
			|\vv_i^{\tranH}\vv_j|\leq\mu,
			\quad i\neq j,
		\end{align*}
		for some $\mu\in[0,1)$.  Let $\mQ_1,\ldots,\mQ_T$ be independent and
		Haar-distributed on $\operatorname{St}(d,p)$.  For every
		$\delta\in(0,1)$, with probability at
		least
		\(
			1-2(d^2-1)
			\exp\!\left(-\frac{\delta^2T}{3(d^2-1)}\right),
		\)
		Assumption~\ref{assump:fisher-frame} holds with
		\begin{align}
			\kappa_F^-
			&=
			e^{-2R^2/\tau}(1-\delta)
			\left(1-\frac1d\right)
			\frac{s_{\mathcal V}}{d^2-1},
			\label{eq:kappa-f-minus}\\
			\kappa_F^+
			&=
			e^{2R^2/\tau}(1+\delta)
			\frac{s_{\mathcal V}}{d^2-1}.
			\label{eq:kappa-f-plus}
		\end{align}
		Moreover,
		\(
			2\left(1-\frac1N\right)(1-\mu^2)
			\leq s_{\mathcal V}\leq2.
		\)
		Thus, when $\mu<1$ is fixed, $s_{\mathcal V}\asymp1$.  Consequently, for fixed $R^2/\tau$,
		$\kappa_F^-\asymp\kappa_F^+\asymp d^{-2}$ once
		$T\gtrsim d^2\log d$.
	\end{lemma}

For $r>0$, define the aligned local tube
\begin{align*}
	\mathcal N_r(\vh)
	:=
	\left\{
	\vh+\vz:
	\vz\in\mathcal H_{\vh},\
	\|\vz\|_2\leq r,\
	\|\vh+\vz\|_2\leq R
	\right\}.
\end{align*}
This set is convex.  Moreover, if $0<r\leq\|\vh\|_2$ and
$\vx=\vh+\vz\in\mathcal N_r(\vh)$, then
\begin{align*}
	\vh^{\tranH}\vx
	\in\R,\quad
	\vh^{\tranH}\vx
	\geq
	\|\vh\|_2\bigl(\|\vh\|_2-r\bigr)
	\geq0.
\end{align*}
Thus $\vx$ is already phase aligned with $\vh$, and
\(
\distT(\vx,\vh)
=
\|\vx-\vh\|_2.
\label{eq:aligned-tube-distance}
\)
\begin{theorem}[Fisher-frame localization from the global slow rate]
		\label{thm:risk-to-distance}
		Suppose Assumptions~\ref{assump:bounded-design} and
		\ref{assump:fisher-frame} hold, and fix $\rho>0$.  There is an absolute
		constant $C>0$ such that, for every $\eta\in(0,1)$, if
		\begin{align}
			T
			\geq
			C
			\frac{R^4\tau^2}
			{(\kappa_F^-)^2\|\vh\|_2^4\rho^4}
			\left\{d+\log\frac1\eta\right\}.
			\label{eq:entry-sample-size}
		\end{align}
		then, with probability at least $1-\eta$,
		\begin{align*}
			\distT(\widehat{\vh}_T,\vh)<\rho.
		\end{align*}
	\end{theorem}

\subsection{Localized parametric excess-risk bound}

	Let
	\begin{align}
		L_H
		:=
		\frac{48R^3}{\tau^3}+\frac{24R}{\tau^2}.
		\label{eq:LH}
	\end{align}

	\begin{theorem}[Sharp local excess-risk and quotient-error bounds]
		\label{thm:local}
		Suppose Assumptions~\ref{assump:bounded-design} and
		\ref{assump:fisher-frame} hold.  Let $\gamma\in(0,1)$ and suppose
		\begin{align}
			\rho
			\leq
			\min\left\{
			\frac{\|\vh\|_2}{2},
			\frac{\kappa_F^-\|\vh\|_2^2}
			{8\tau^2L_H}
			\right\}.
			\label{eq:local-radius-condition}
		\end{align}
		There is an absolute constant $C>0$ such that, if
		\begin{align}
			T
			\geq
			C\max\left\{
			\frac{R^4\tau^2}
			{(\kappa_F^-)^2\|\vh\|_2^4\rho^4}
			\left(d+\log\frac3\gamma\right),
			\frac{\tau^2}{(\kappa_F^-)^2\|\vh\|_2^4}
			\log\frac{12d}{\gamma},
			\frac{\tau}{\kappa_F^-\|\vh\|_2^2}
			\log\frac{12d}{\gamma},
			\frac{R^2}{\kappa_F^-\|\vh\|_2^2}
			\left(d+\log\frac6\gamma\right)
			\right\},
			\label{eq:local-sample-size}
		\end{align}
		then, with probability at least $1-\gamma$, one has
		\begin{align}
			\overbar L_T(\widehat{\vh}_T)-\overbar L_T(\vh)
			&\leq
			C\frac{d+\log(1/\gamma)}{T},
			\label{eq:local-excess-fisher}\\
			\distT^2(\widehat{\vh}_T,\vh)
			&\leq
			C\frac{\tau^2\{d+\log(1/\gamma)\}}
			{T\kappa_F^-\|\vh\|_2^2}.
			\label{eq:local-parameter-frame}
		\end{align}
	\end{theorem}

\subsection{Local minimax lower bound}
\label{sec:minimax}

	Fix $\vh_0\in\C^d\setminus\{\bm0\}$ and write
	\begin{align*}
		r_0:=\|\vh_0\|_2,
		\qquad
		r_+:=r_0+\rho.
	\end{align*}
	Define
	\begin{align*}
		\Theta(\vh_0,\rho)
		:=
		\{\vh\in\C^d:\|\vh-\vh_0\|_2\leq\rho\}.
	\end{align*}
	Let $P_{\vh}^{(T)}$ be the joint law of
	$I^{(T)}=(I_1,\ldots,I_T)$ under channel $\vh$, and define
	\begin{align*}
		\mathcal R_T(\vh_0,\rho)
		:=
		\inf_{\widehat{\vh}}
		\sup_{\vh\in\Theta(\vh_0,\rho)}
		\E[\vh]{\distT^2(\widehat{\vh},\vh)}.
	\end{align*}
	
	\begin{theorem}[Fisher-aware complex local minimax lower bound]
		\label{thm:minimax-lower}
		Suppose $d\geq9$, $T\geq1$, $\tau>0$,
		$0<\rho\leq r_0/4$, and $r_+\leq R$.  Assume conditional independence,
		the Gumbel--Softmax model \eqref{eq:intro-observation-model}, and the upper
		inequality in Assumption~\ref{assump:fisher-frame} with constant
		$\kappa_F^+$.  Then
		\begin{align}
			\mathcal R_T(\vh_0,\rho)
			\geq
			c\min\left\{
			\rho^2,
			\frac{\tau^2d}{T\kappa_F^+r_+^2}
			\right\},
			\label{eq:minimax-fisher-lower}
		\end{align}
		where $c>0$ is an absolute constant.
	\end{theorem}
	
	\begin{remark}
		\label{rem:minimax-regimes}
		The lower bound has two regimes.  When
		\begin{align*}
			T\lesssim
			\frac{\tau^2d}{\kappa_F^+\rho^2r_+^2};
		\end{align*}
		the radius term is active, and the observations do not contain enough
		information to localize the channel within the prescribed neighborhood.
		When
		\begin{align}
			T\gtrsim
			\frac{\tau^2d}{\kappa_F^+\rho^2r_+^2},
			\label{information-region}
		\end{align}
		the information term is active.  Since $r_0\leq r_+\leq5r_0/4$, this term
		is of order $\tau^2d/(T\kappa_F^+\|\vh_0\|_2^2)$.
	\end{remark}
	
	\begin{remark}
		\label{rem:upper-lower-comparison}
		Suppose that
		the radius and sample-size conditions of Theorem~\ref{thm:local} hold, and that
		the lower bound in Theorem~\ref{thm:minimax-lower} is in the information
		regime \eqref{information-region}, and that $\log(1/\gamma)=O(d)$.  Since
		$r_0\leq r_+\leq5r_0/4$, the two bounds have the respective forms
		\begin{align*}
			\distT^2(\widehat{\vh}_T,\vh_0)
			&\lesssim
			\frac{\tau^2d}{T\kappa_F^-r_0^2}, \quad \text{with probability at least $1-\gamma$}\\
			\mathcal R_T(\vh_0,\rho)
			&\gtrsim
			\frac{\tau^2d}{T\kappa_F^+r_0^2}.
		\end{align*}
		Thus they have the same explicit dependence on $T$, $d$, $\tau$, and the
		channel norm, up to the design-dependent ratio
		$\kappa_F^+/\kappa_F^-$ and absolute constants.
		For the Haar--Stiefel design in
		Lemma~\ref{lem:complex-secant-stiefel}, on the high-probability design
		event stated there, fixed $R^2/\tau$ and $\delta$ bounded away from one
		give
		$\kappa_F^-\asymp\kappa_F^+\asymp s_{\mathcal V}/d^2$.
		Consequently, under the regimes specified above, the upper and lower
		bounds both have scale
		\begin{align*}
			\frac{\tau^2d^3}{T r_0^2s_{\mathcal V}}
			\asymp
			\frac{\tau^2d^3}{T r_0^2},
		\end{align*}
		where the final equivalence uses $s_{\mathcal V}\asymp1$ for the fixed
		nondegenerate codebook specified above.
	\end{remark}

%%%%%%%%%%%%%%%%%%%%%%%%

%%%%%%%%%%%%%%%%%%%%%%%%
\section{Experiments on FDD Channels}\label{sec:numerics}
\subsection{Experimental Setup}
\begin{table}[ht!]
\centering
\caption{Key parameters for QuaDRiGa-based channel generation.}
\label{tab:fdd_data}
\begin{tabular}{ll}
\toprule
\textbf{Parameter} & \textbf{Value} \\
\midrule
Propagation scenario & 3GPP 3D UMa \\
Carrier frequency & 1.84 GHz (DL), 1.74 GHz (UL) \\
BS antenna array & $8 \times 2$ cross-polarized array, 32 antenna elements\\
BS antenna spacing & $0.5\lambda$ (horizontal), $0.7\lambda$ (vertical) \\
BS downtilt angle & $7^\circ$ \\
UE antenna configuration & Cross-polarized, $4$ antennas \\
BS height & 30 m \\
UE height & 1.5 m \\
Cell/user distribution range & 300 m radius \\
Minimum BS--UE distance & 35 m \\
Subcarrier spacing & 180 kHz \\
System bandwidth & 10 MHz \\
\bottomrule
\end{tabular}
\end{table}

Here we assess the practical applicability of the proposed MLE by applying it to PMI observations generated according to the deterministic hard-PMI rule used in FDD systems. The underlying FDD channel data are generated using QuaDRiGa \cite{Jaeckel2014QuaDRiGa}. The dataset is constructed under a three-dimensional urban macro-cell (3D UMa) scenario, and the detailed simulation parameters are summarized in Table~\ref{tab:fdd_data}. We generate channel realizations for 50 users. For each user, the channel is sampled over 50 subcarriers and 400 time snapshots, all uniformly selected. For evaluation, we retain 4 time snapshots and 2 subcarriers for each user, yielding a total of 400 test samples. We consider the case where  the base station has $d = 32$ transmit antennas in our experiments.
In accordance with the effective CSI feedback configuration in ~\cite{3gpp38214v1620_codebook_pmi},  the reduced dimension is set to $p=8$. Thus,  the user observes an equivalent eight-port channel after dimensionality-reduction.

It is worth pointing out again that the observation index in the numerical experiments is drawn from the hard PMI model \eqref{eq:pmi}, while the CQI is given by \eqref{eq:CQI}.
Moreover, as in~\cite{luoFDD}, the CQI is further quantized using $32$ bits.
The reconstruction quality is evaluated in terms of the beam-precision metric:
\[
    \mathrm{Beam\,Precision}
    =
    \frac{
\widehat{\vu}^{\tranH}\vh\vh^\tranH\widehat{\vu}
    }
    {
    \|\vh\|_2^2
   },
\]
where $\widehat{\vu}=\widehat{\vh}/\|\widehat{\vh}\|_2$ is the direction of the reconstructed channel. This metric measures how well the reconstructed beam space aligns with the direction of the true channel. In particular, it emphasizes directional consistency and  is independent of the power scaling.

\subsection{Baseline Methods}

The proposed MLE method is compared against three baselines: spectral method, alternating minimization (AM), and subspace phase retrieval (Subspace-PR). 

\paragraph{Spectral method.}
The spectral method is introduced by Yang~\cite{CN105959046A}, which estimates the beam direction by the principal direction of
\[
    \widehat{\mH}_{\mathrm{spec}}
    =
    \frac{1}{T}\sum_{t=1}^T \mQ_t \vv_{I_t}\vv_{I_t}^\tranH \mQ_t^\tranH.
    \label{eq:spectral_fdd}
\]
In particular, when \(T=1\), the spectral method reduces to the two-stage precoding baseline~\cite{xu2014user,liu2015two},
\begin{align*}
\widehat{\vh}_{\mathrm{Type-I}}=\mQ_1\vv_{I_1},\numberthis\label{eq:tw-stage precoding}
\end{align*}
where the dimensionality-reduction matrix \(\mQ_1\) acts as the outer precoder and \(\vv_{I_1}\) is the Type-I codebook~\cite{3gpp38214v1620_codebook_pmi} entry indexed by the PMI.

% In particular, when $T=1$, the spectral method reduces to the Type-I baseline \cite{3gpp38802},
% \[
% \widehat{\mH}_{\mathrm{Type-I}}=\mQ_1\mV_{I_1}.
% \]
%which corresponds to the ``Type-I Codebook Line'' in the figures.

\paragraph{Alternating minimization (AM).}
The alternating minimization method is introduced in~\cite{luoFDD}, where the estimate is obtained as a solution to the following optimization problem that incorporates the CQI values:
\[
    \widehat{\vh}_{\mathrm{AM}}
    \in
    \argmin_{\vx\in\C^{d}}
    \sum_{t=1}^T
    \left(
    \left|
    \vv_{I_t}^\tranH \mQ_t^\tranH \vx
    \right|
    -
    \sqrt{\eta_t}
    \right)^2
    +
    \lambda_{\mathrm{AM}}\|\vx\|_2^2.
    \label{eq:am_single}
\]
By introducing auxiliary phases $\{\phi_t\}_{t=1}^T$, this problem can be reformulated as
\[
    \widehat{\vh}_{\mathrm{AM}}
    \in
    \argmin_{\vx,\{\phi_t\}}
    \sum_{t=1}^T
    \left|
    \vv_{I_t}^\tranH \mQ_t^\tranH \vx\, e^{j\phi_t}
    -
    \sqrt{\eta_t}
    \right|^2
    +
    \lambda_{\mathrm{AM}}\|\vx\|_2^2,
    \label{eq:am_single_phase}
\]
and can be solved via alternating minimization.

\paragraph{Subspace phase retrieval (Subspace-PR).}
Downlink reconstruction from PMI and CQI can be viewed as a quadratic inverse problem, or equivalently, a phase retrieval problem \cite{fannjiang2020numerics}. Furthermore, although uplink and downlink channels are not instantaneously reciprocal in FDD systems, they typically share partially common spatial structure under the same propagation environment~\cite{fddhuawei_Reciprocity,yin2022partial}. This motivates the subspace phase retrieval (Subspace-PR) method introduced in~\cite{li2023csi_subpr}.
More precisely, let $\mB_k=\mathrm{eigvecs}(\bm{\Sigma}_{\mathrm{ul}},k)\in\C^{d\times k}$ be a basis for the dominant uplink subspace. Then the  downlink beam matrix is then parameterized as
\[
\widehat{\mH}_{\mathrm{subspace\mbox{-}pr}}=\mB_k \vs,
\quad
\vs\in\C^{k\times 1},
\]
which reduces the number of unknown parameters from the ambient antenna dimension $d$ to the low-dimensional coefficient matrix $\mS$. We consider the following two subspace phase-retrieval formulations:
\[
    \widehat{\vs}_{\mathrm{subspace\mbox{-}pr}}
    \in
    \argmin_{\vs\in\C^{k\times 1}}
    \frac{1}{T}\sum_{t=1}^T
    \left|
    \|
    \vv_{I_t}^\tranH
    \mQ_t^\tranH
    \mB_k \vs
    \|_F^2
    -
    \eta_t
    \right|^2,
    \label{eq:subspace_pr_wf}
\]
and
\[
    \widehat{\vs}_{\mathrm{subspace\mbox{-}pr}}
    \in
    \argmin_{\vs\in\C^{k\times 1}}
    \frac{1}{T}\sum_{t=1}^T
    \left|
    \|
    \vv_{I_t}^\tranH
    \mQ_t^\tranH
    \mB_k \vs
    \|_F
    -
    \sqrt{\eta_t}
    \right|^2.
    \label{eq:subspace_pr_af}
\]
The first formulation is solved via Wirtinger Flow (WF) \cite{candes2015phase}, and the second via Amplitude Flow (AF) \cite{wang2018taf}. We report the better of the two results for each test case.

Note that, as with the subspace phase-retrieval method, we  also incorporate an uplink-informed subspace prior into the MLE, leading to the subspace-constrained formulation
\begin{equation}
    \widehat{\vs}
    \in
    \argmin_{\vs\in\C^{k\times 1}}
    \frac{1}{T}\sum_{t=1}^T
    \log
    \sum_{j=1}^N
    \exp\!\left(
    \frac{
    \| \vv_{j}^\tranH \mQ_t^\tranH \mB_k\vs \|_F^2
    -
    \| \vv_{I_t}^\tranH \mQ_t^\tranH \mB_k\vs \|_F^2
    }{\tau}
    \right),
    \label{eq:fdd_subspace_mle}
\end{equation}
followed by
\[
\widehat{\vh}_{\mathrm{MLE}}=\mB_k \widehat{\vs},
\quad
\widehat{\vs}\in\C^{k\times 1}.
\]
It is evident that, if the ground-truth channel satisfies $\vh\in\mathrm{span}(\mB_k)$ with some coefficient matrix $\vs$, then the estimation problem is exactly equivalent to the original one but restricted to a known $k$-dimensional subspace. Therefore, the same theoretical analysis applies to the coefficient variable $\vs$ after replacing the ambient dimension $d$ with the effective subspace dimension $k$.

\subsection{Implementation Details}
We adopt gradient descent to find the solution of the MLE, with the maximum number of iterations set to $100$.  The same  iteration budget is also used in AM and the gradient-based methods for Subspace-PR. In addition, the algorithms are also terminated whenever the two consecutive iterates are sufficiently close to each other (relative change less than $10^{-3}$), where relative change is evaluated as
\[
\frac{\bigl\|\bm{x}^{(t)}-e^{-j\phi_t}\bm{x}^{(t-1)}\bigr\|_2}{\bigl\|\bm{x}^{(t-1)}\bigr\|_2},
\quad
\phi_t=\arg\left((\bm{x}^{(t-1)})^{\mathsf H}\bm{x}^{(t)}\right).
\]
While the original paper adopts random initialization for AM, we find in our experiments that initializing AM with the spectral estimate  yields better performance, which is used in our experiments.  As observed in \cite{luoFDD}, the regularization parameter $\lambda_{\mathrm{AM}}$ has a substantial impact on performance. We thus tune $\lambda_{\mathrm{AM}}$ on a validation set constructed from uplink channel samples available at the base station, using the grid
\[
\{0.001,\,0.01,\,0.1,\,1,\,10,\,100,\,1000\}.
\]
Based on this validation, the selected value is $\lambda_{\mathrm{AM}}=1$.
For Subspace-PR, we use a subspace-aware spectral initialization. Specifically, we form the reduced-domain sample covariance
\[
\mC_{\mathrm{spec}}
=
\frac{1}{T}\sum_{t=1}^T
(\mB_k^\tranH \mQ_t)\vv_{I_t}\vv_{I_t}^\tranH(\mB_k^\tranH \mQ_t)^\tranH,
\]
and set
\(
\vs_{\mathrm{init}}^{\mathrm{spec}}
=
\mathrm{eigvecs}(\mC_{\mathrm{spec}},1).
\)

As the ablation study indicates, the performance of the MLE is insensitive to the choice of initialization. Hence, we adopt the first  column of an identity matrix as the default in all experiments. For both the Subspace-PR and the MLE, the subspace dimension is set to $k = 8$ (i.e., $\mB_k \in \mathbb{C}^{d \times 8}$).  In addition, we set $\tau=1$ 
for the MLE, which is a natural default and performs reasonably well in the ablation study.
% In addition, we set $\tau = 1$ for the MLE, which yields the best performance in the ablation study.

 When $T=1$, we use a Type-I compatible dimensionality-reduction matrix,
\[
    \mQ_1
    =
    \mathrm{eigvecs}(\Sigma_{\mathrm{ul}},8)
    \left(
    \mI_2 \otimes
    \frac{\mathrm{DFT}(2)\otimes \mathrm{DFT}(2)}{2}
    \right),
    \label{eq:q_type1}
\]
which combines the user-specific uplink subspace information with the dual-polarized Type-I codebook structure~\cite{3gpp38214v1620_codebook_pmi}. When $T>1$, diversity across the matrices $\mQ_t$ becomes essential. We therefore consider the following two options (note that the first dimensionality-reduction matrix is always set to $\mQ_1$):
\begin{itemize}
    \item Fully random design: each $\mQ_t\in\C^{32\times 8}$ is generated independently from a complex random Gaussian matrix, followed by orthogonalization.

    \item Structured design with fixed outer and random inner layers:
\begin{equation}
    \mQ_t=\mQ_{\mathrm{out}}\mU_t,
    \label{eq:q_outer_inner}
\end{equation}
where $\mQ_{\mathrm{out}} = \mathrm{eigvecs}(\bm{\Sigma}_{\mathrm{ul}},8) \in\C^{32\times 8}$ is  fixed, and $\mU_t\in\C^{8\times 8}$ is a random unitary matrix generated independently at each round. This construction preserves the dominant user subspace in the outer layer while injecting additional diversity through the random inner layer.
\end{itemize}

\subsection{Results}

\begin{figure}[ht!]
    \centering
    \begin{minipage}[b]{0.45\linewidth}
        \centering
        \includegraphics[width=\linewidth]{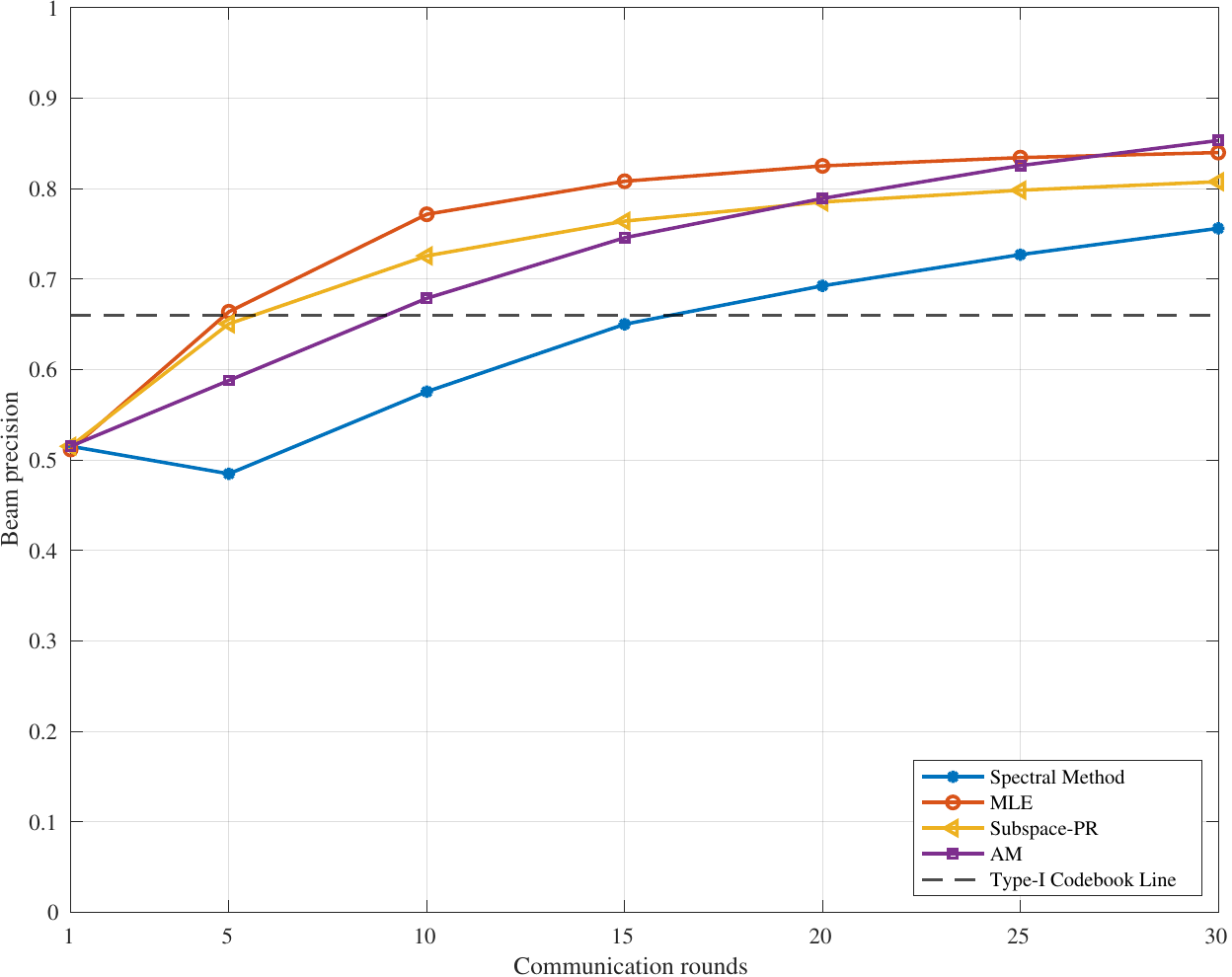}
        \label{fig:fdd_randQ_r1}
    \end{minipage}
    \hfill
    \begin{minipage}[b]{0.45\linewidth}
        \centering
        \includegraphics[width=\linewidth]{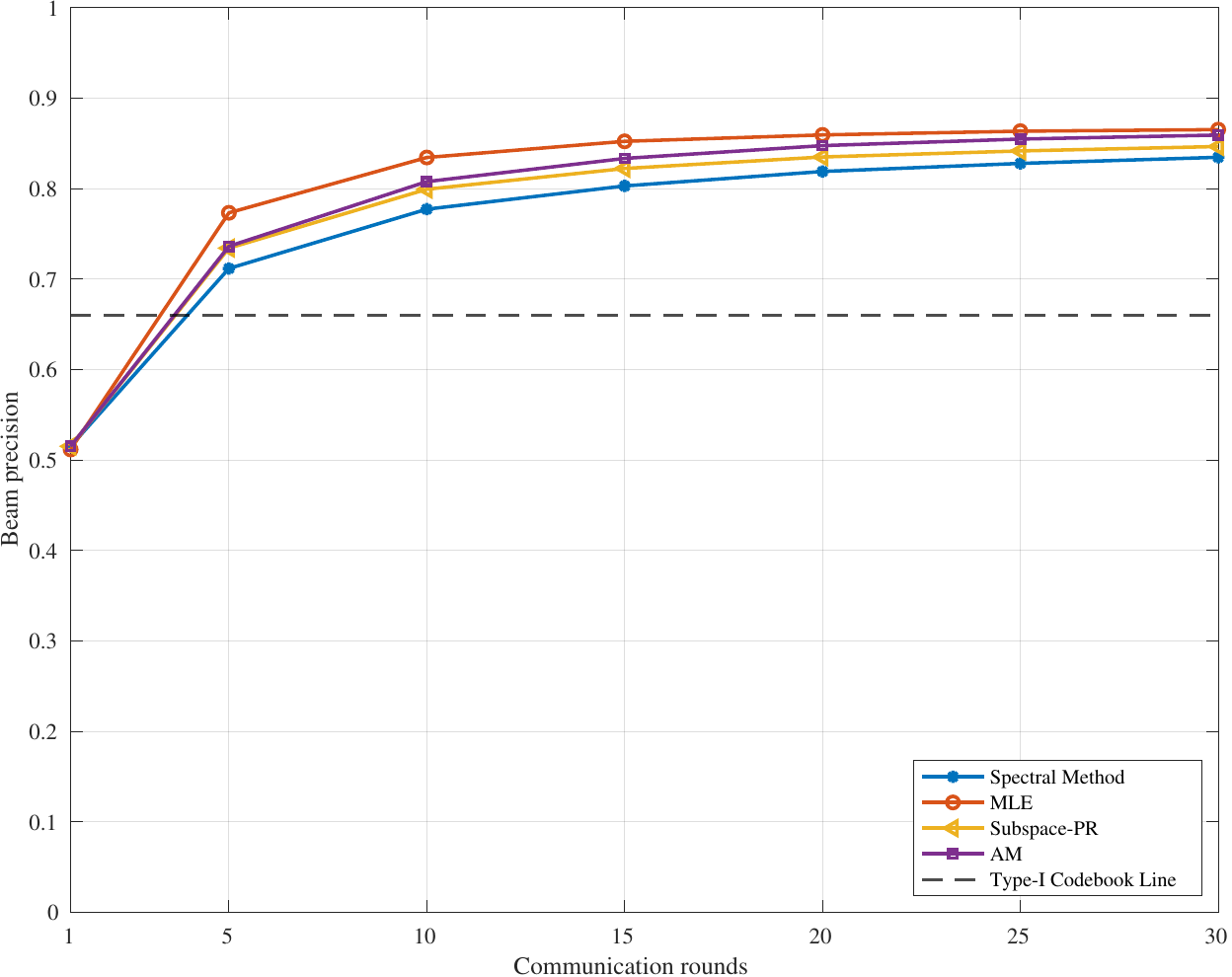}
        \label{fig:fdd_QcovQin_r1}
    \end{minipage}
    \caption{Beam precision versus communication rounds. Left: fully random dimensionality-reduction matrices; Right: structured dimensionality-reduction matrices. The dashed horizontal line corresponds to the two-stage precoding baseline in \eqref{eq:tw-stage precoding}.}
    \label{fig:fdd_r1}
\end{figure}

\paragraph{Comparison of different methods.}
Figures~\ref{fig:fdd_r1}  reports the beam precision versus the number of communication rounds for the aforementioned methods. 
Overall, the proposed MLE achieves a superior performance under both constructions of $\mQ_t$. Under fully random $\mQ_t$, the spectral method exhibits high variance for small $T$ and can even perform worse than the Type-I codebook baseline, whereas the proposed MLE improves steadily as $T$ grows and consistently outperforms both AM and Subspace-PR. Under the structured design $\mQ_t=\mQ_{\mathrm{out}}\mU_t$, all methods improve considerably, indicating that incorporating the uplink-informed outer transform makes each feedback round more informative. Even in this more favorable setting, however, the proposed MLE remains the strongest method across all tested communication rounds.
Therefore, the proposed MLE  method is highly competitive even though it is driven soly by  PMI feedback, while the baseline methods such as AM and Subspace-PR additionally rely on CQI. This suggests that the likelihood-based formulation effectively captures the comparative information conveyed by the PMI feedback.

\paragraph{Ablation study on $\tau$  and initialization for MLE.}
\begin{figure}[ht!]
    \centering
    \begin{minipage}[b]{0.45\linewidth}
        \centering
        \includegraphics[width=\linewidth]{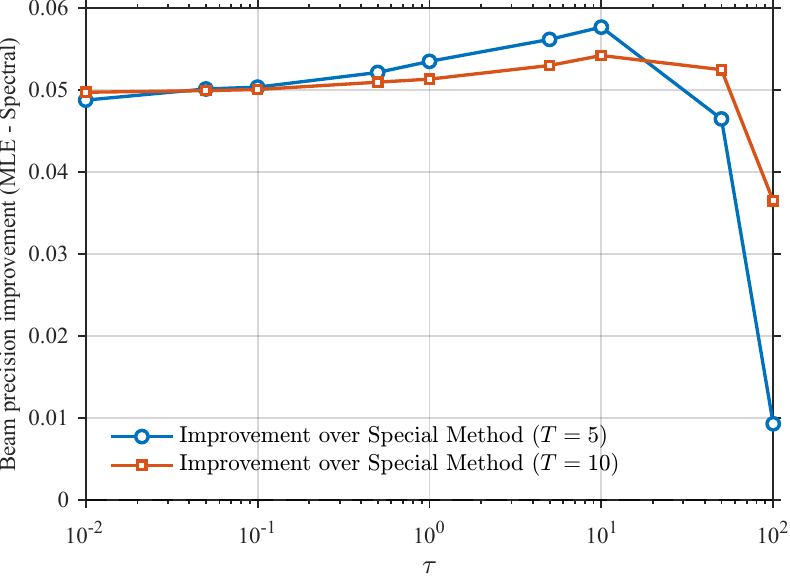}
    \end{minipage}
    \hfill
    \begin{minipage}[b]{0.45\linewidth}
        \centering
        \includegraphics[width=\linewidth]{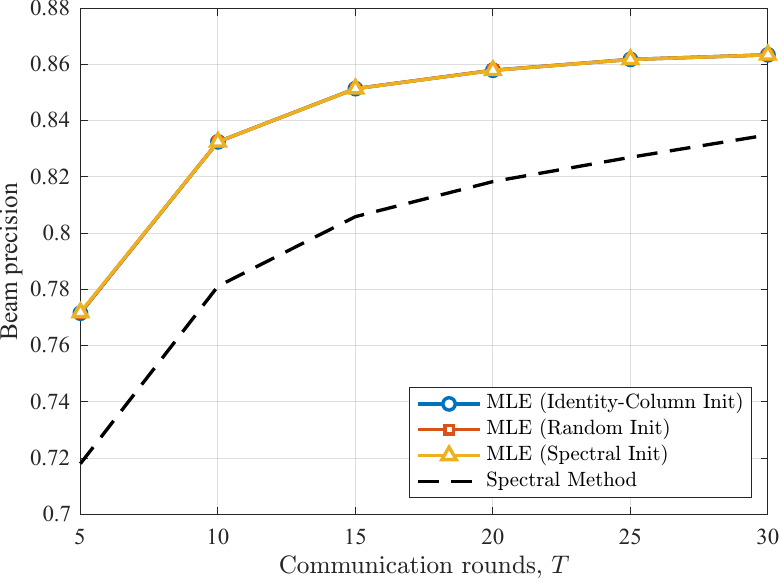}
    \end{minipage}
    \caption{Left: Ablation study on the temperature parameter $\tau$  under the structured dimensionality-reduction design, where the curves show the beam-precision improvement of the proposed MLE over the spectral method for $T=5$ and $T=10$; Right: Ablation study on initialization  under the structured dimensionality-reduction design, where identity-column initialization, random initialization, and spectral initialization are tested.}
    \label{fig:tau_ablation}
\end{figure}

We next investigate the role of the temperature parameter $\tau$ in the MLE based on the structured dimensionality-reduction design \eqref{eq:q_outer_inner}. Figure~\ref{fig:tau_ablation} (left) shows the beam-precision improvement of the proposed MLE over the spectral method.
It can be seen that, the proposed MLE is relatively insensitive to the choice of $\tau$ over a broad range of values. For both $T=5$ and $T=10$, the improvement over the spectral method remains positive throughout the entire range of tested values, with only mild variation as $\tau$ increases from small to moderately large values. The best performance is achieved at an intermediate-to-large temperature, beyond which the gain begins to decline as $\tau$ becomes excessively large. Nonetheless, the MLE still retains a clear advantage over the spectral method.

We also compare three initialization strategies for the low-dimensional coefficient matrix $\vs\in\C^{k\times 1}$ in the subspace parameterization $\widehat{\vh}=\mB_k\vs$. In addition to the spectral initialization and the initialization based on the first  column of an identity matrix, we also consider 
the {random initialization}, where we draw a complex random  Gaussian vector, followed by normalization. 
As Figure~\ref{fig:tau_ablation} (right) shows, the proposed MLE is largely insensitive to the choice of initialization, and three initialization schemes yield similar performance and consistently outperform the spectral method. Therefore, we use the identity-column initialization as the default choice in the earlier experiments.\\

%%%%%%%%%%%%%%%%%%%%%%%%
\section{Proofs of Main Results}
\label{sec:proofs-main}

\subsection{Auxiliary Lemmas}
We first state several auxiliary lemmas used in the main arguments,
in the order in which they arise.

\begin{lemma}
\label{lemma-differential}
For $\vx,\vz\in\C^d$, define
\begin{align*}
	\mC_t(\vx)
	&:=
	\sum_{i=1}^Np_t(i;\vx)\mA_{t,i},\\
	\mS_t(\vx)[\vz]
	&:=
	\sum_{i=1}^Np_t(i;\vx)
	\innerR{\mA_{t,i}\vx}{\vz}\mA_{t,i}\vx,\\
	\vy_t(\vx)
	&:=
	\sum_{i=1}^Np_t(i;\vx)\mA_{t,i}\vx
	=
	\mC_t(\vx)\vx.
\end{align*}
Then one has
\begin{align}
	\nabla\ell_t(i;\vx)&=	\frac{2}{\tau}	(\mC_t(\vx)-\mA_{t,i})\vx,	\label{eq:sample-gradient}\\
	\nabla L_T(\vx) &= \frac{2}{\tau T} \sum_{t=1}^T (\mC_t(\vx)-\mA_{t,I_t})\vx,\\
	\nabla \overbar{L}_T(\vx) &=\frac{2}{\tau T}	\sum_{t=1}^T	(\mC_t(\vx)-\mC_t(\vh))\vx,\\
	\nabla^2 L_T(\vx)[\vz] &=
	-\frac2{\tau T}\sum_{t=1}^T\mA_{t,I_t}\vz
	+\frac2{\tau T}\sum_{t=1}^T\mC_t(\vx)\vz
	+\frac4{\tau^2T}\sum_{t=1}^T
	\left(
	\mS_t(\vx)[\vz]
	-\innerR{\vy_t(\vx)}{\vz}\vy_t(\vx)
	\right),	\label{eq:sample-hessian}\\
	\nabla^2 \overbar{L}_T(\vx)[\vz] &=-\frac2{\tau T}\sum_{t=1}^T\mC_t(\vh)\vz
	+\frac2{\tau T}\sum_{t=1}^T\mC_t(\vx)\vz
	+\frac4{\tau^2T}\sum_{t=1}^T
	\left(
	\mS_t(\vx)[\vz]
	-\innerR{\vy_t(\vx)}{\vz}\vy_t(\vx)
	\right),
	\label{eq:population-hessian}\\
	\innerR{\nabla^2\overbar L_T(\vh)[\vz]}{\vz} &=\frac1{2\tau^2T}
	\sum_{t=1}^T\sum_{i=1}^N\sum_{j=1}^N
	p_t(i;\vh)p_t(j;\vh)
	\innerR{\mA_{t,i}-\mA_{t,j}}
	{\vz\vh^{\tranH}+\vh\vz^{\tranH}}^2.	\label{eq:population-hessian-pairwise}
\end{align}
\end{lemma}

\begin{lemma}[Phase Procrustes alignment]
\label{lem:phase-procrustes-alignment}
For every $\vx\in\C^d$, let 
$\phi_{\vx}\in[0,2\pi)$ be a minimizing rotation angle in the definition of $\distT(\vx,\vh)$. Then the
aligned representative
\(
	\vx^{\mathrm a}
	:=
	e^{-\mathrm i\phi_{\vx}}\vx
\)
satisfies
\begin{align*}
	\vh^{\tranH}\vx^{\mathrm a}
	=
	|\vh^{\tranH}\vx|
	\geq0,\quad
	\vx^{\mathrm a}-\vh
	\in
	\mathcal H_{\vh},\quad
	\|\vx^{\mathrm a}-\vh\|_2
	=
	\distT(\vx,\vh).
\end{align*}
%If $\vh^{\tranH}\vx\neq0$, then
%$\phi_{\vx}\equiv\arg(\vh^{\tranH}\vx)\pmod{2\pi}$.
\end{lemma}

For each $t$, define the following real function on the real Hilbert space of Hermitian matrices,
\begin{align*}
	f_t(\mM)
	:=
	\log\sum_{j=1}^N
	\exp\!\left(\innerR{\mA_{t,j}}{\mM}/\tau\right).
\end{align*}
%The following differential identities of \(f_t(\mM)\) will be used.  
\begin{lemma}[Lifted log-partition derivatives]
\label{lem:lifted-log-partition}
For every $t$, every Hermitian matrix $\mM$, and every Hermitian direction
$\mU$,
\begin{align*}
	\nabla f_t(\mM)
	&=
	\frac1\tau
	\sum_{i=1}^Nq_t(i;\mM)\mA_{t,i},\\
	\nabla^2f_t(\mM)[\mU]
	&=
	\frac1{2\tau^2}
	\sum_{i=1}^N\sum_{j=1}^N
	q_t(i;\mM)q_t(j;\mM)
	\innerR{\mA_{t,i}-\mA_{t,j}}{\mU}
	(\mA_{t,i}-\mA_{t,j}).
\end{align*}
Consequently,
\begin{align}
	\innerR{\nabla^2f_t(\mM)[\mU]}{\mU}
	=
	\frac1{2\tau^2}
	\sum_{i=1}^N\sum_{j=1}^N
	q_t(i;\mM)q_t(j;\mM)
	\innerR{\mA_{t,i}-\mA_{t,j}}{\mU}^2.
	\label{eq:lifted-hessian}
\end{align}
\end{lemma}

 	\begin{lemma}[Weighted-risk separation under the Fisher frame]
 		\label{lem:fisher-frame-consequences}
 		Under Assumption~\ref{assump:fisher-frame}, for every
 		$\|\vx\|_2\leq R$,
 		\begin{align}
 			\frac{\kappa_F^-}{4\tau^2}
 			\fronorm{\vx\vx^{\tranH}-\vh\vh^{\tranH}}^2
 			&\leq
 			\overbar L_T(\vx)-\overbar L_T(\vh)
 			\leq
 			\frac{\kappa_F^+}{4\tau^2}
 			\fronorm{\vx\vx^{\tranH}-\vh\vh^{\tranH}}^2.
 			\label{eq:weighted-risk-sandwich}
 		\end{align}
 	\end{lemma}

 \begin{lemma}[Rank-one quotient geometry]
 	\label{lem:rank-one-quotient}
 	For every $\vx,\vh\in\C^d$,
 	\begin{align}
 		\fronorm{\vx\vx^{\tranH}-\vh\vh^{\tranH}}^2
 		\geq
 		\frac12\|\vh\|_2^2\distT^2(\vx,\vh).
 		\label{eq:rank-one-quotient}
 	\end{align}
 \end{lemma}

	\begin{lemma}[Horizontal Fisher geometry under the weighted frame]
 		\label{lem:horizontal-fisher-frame}
 		Suppose Assumption~\ref{assump:fisher-frame} holds and
 		$0<\|\vh\|_2\leq R$.  Define the horizontal Fisher information operator
 		$\mathcal I_{\vh}:\mathcal H_{\vh}\to\mathcal H_{\vh}$ by
 		\begin{align}
 			\innerR{\mathcal I_{\vh}\vz}{\vw}
 			:=
 			\innerR{\nabla^2\overbar L_T(\vh)[\vz]}{\vw},
 			\qquad
 			\vz,\vw\in\mathcal H_{\vh}.
 			\label{eq:horizontal-fisher}
 		\end{align}
 		This operator is self-adjoint.  Let
 		\begin{align*}
 			\lambda_-
 			:=
 			\lambda_{\min}(\mathcal I_{\vh}),
 			\quad
 			\lambda_+
 			:=
 			\lambda_{\max}(\mathcal I_{\vh}).
 		\end{align*}
 		Then
 		\begin{align}
 			\lambda_-
 			\geq
 			\frac{\kappa_F^-\|\vh\|_2^2}{\tau^2},
 			\quad
 			\lambda_+
 			\leq
 			\frac{2\kappa_F^+\|\vh\|_2^2}{\tau^2}.
 			\label{eq:fisher-eigen-frame}
 		\end{align}
 		In particular, $\mathcal I_{\vh}$ is positive definite on
 		$\mathcal H_{\vh}$, and
 		\begin{align*}
 			\|\vz\|_{\mathcal I_{\vh}}^2
 			:=
 			\innerR{\mathcal I_{\vh}\vz}{\vz}
 		\end{align*}
 		defines a norm on $\mathcal H_{\vh}$.
 	\end{lemma}
 	
 	It is easy to see that the sample gradients at the true parameter $\vh$ lie in the horizontal space, and so does \(\nabla L_T(\vh)\).
 	Whenever $\lambda_->0$, the inverse square root
 	$\mathcal I_{\vh}^{-1/2}:\mathcal H_{\vh}\to\mathcal H_{\vh}$
 	is therefore well-defined on each sample gradient and on
 	$\nabla L_T(\vh)$.

    \begin{lemma}[Whitened-score concentration]
 		\label{lem:score-concentration}
 		Suppose Assumption~\ref{assump:bounded-design} holds and
 		$I_1,\ldots,I_T$ are independent observations from
 		\eqref{eq:intro-observation-model}.  Let $\mathcal I_{\vh}$ be defined
 		by \eqref{eq:horizontal-fisher}, and suppose
 		$\lambda_-:=\lambda_{\min}(\mathcal I_{\vh})>0$.
 		There is an absolute constant $C>0$ such that, for every
 		$\delta\in(0,1)$, if
 		\begin{align*}
 			T
 			\geq
 			C\frac{R^2}{\tau^2\lambda_-}
 			\left(d+\log\frac2\delta\right),
 		\end{align*}
 		then, with probability at least $1-\delta$,
 		\begin{align}
 			\left\|
\bm I_{\bm h}^{-1/2}
\nabla L_T(\bm h)
\right\|_2^2
 			\leq
 			C\frac{d+\log(2/\delta)}{T}.
 			\label{eq:whitened-score-concentration}
 		\end{align}
 		Here $\mathcal I_{\vh}^{-1/2}$ acts on $\mathcal H_{\vh}$, and
 		the probability is over the observations conditional on the design.
 	\end{lemma}

 	\begin{lemma}[Empirical curvature relative to Fisher information]
 		\label{lem:local-rsc}
 		Suppose $\lambda_->0$,
 		\begin{align*}
 			\rho
 			\leq
 			\frac{\lambda_-}
 			{8L_H},
 		\end{align*}
 		and
 		\begin{align*}
 			T
 			\geq
 			C_{\mathrm{rsc}}\max\left\{
 			\frac{\log(4d/\delta)}{\tau^2\lambda_-^2},
 			\frac{\log(4d/\delta)}{\tau\lambda_-}
 			\right\}.
 		\end{align*}
 		Then, with probability at least $1-\delta$,
 		\begin{align}
 			\innerR{\nabla^2L_T(\vx)[\vz]}{\vz}
 			\geq
 			\frac34\innerR{\mathcal I_{\vh}\vz}{\vz},
 			\qquad
 			\vx\in\mathcal N_\rho(\vh),\quad
 			\vz\in\mathcal H_{\vh}.
 			\label{eq:empirical-horizontal-curvature}
 		\end{align}
 	\end{lemma}

% \begin{lemma}[Mutual-information bound]
% \label{lem:mi-bound}
% Let $m:=2d-1$, let $\vb_1,\ldots,\vb_m$ be any real orthonormal basis of
% the horizontal space $\mathcal H_{\vh_0}:= \left\{
% 	\vz\in\C^d:
% 	\innerR{\mathrm i\vh_0}{\vz}=0
% 	\right\}$, and let
% $\Omega\subseteq\{0,1\}^m$ be finite.  For $a>0$, define
% \begin{align*}
% 	\vz_\omega
% 	:=
% 	a\sum_{j=1}^m\omega_j\vb_j,
% 	\quad
% 	\vh_\omega
% 	:=
% 	\vh_0+\vz_\omega,
% 	\quad \omega\in\Omega.
% \end{align*}
% Suppose that $\max_{\omega\in\Omega}\|\vz_\omega\|_2\leq	\rho$. If $W$ is uniform on $\Omega$ and, conditional on $W=\omega$, the data
% have law $P_{\vh_\omega}^{(T)}$, then, under
% $\|\va_{t,i}\|_2\leq1$,
% \begin{align*}
% 	I(W;I^{(T)})
% 	\leq
% 	\frac{Tr_+^2}{2\tau^2|\Omega|}
% 	\sum_{\omega\in\Omega}
% 	\|\vh_\omega-\vh_0\|_2^2
% 	\leq
% 	\frac{Tr_+^2a^2m}{2\tau^2}.
% \end{align*}
% \end{lemma}

%\subsection{Proofs of the main theorems}

\subsection{Proof of Theorem~\ref{thm:global}}
Let $Z_T$ be defined as
\begin{align*}
	Z_T
	:=
	\sup_{\|\vx\|_2\leq R}
	\left\{
	\overbar L_T(\vx)-\overbar L_T(\vh)
	-L_T(\vx)+L_T(\vh)
	\right\}.
\end{align*}
Because $\vh$ is feasible and $\widehat{\vh}_T$ minimizes the empirical loss, one has 
\begin{align*}
	\overbar L_T(\widehat{\vh}_T)-\overbar L_T(\vh)
	&=
	\overbar L_T(\widehat{\vh}_T)-L_T(\widehat{\vh}_T)
	+L_T(\widehat{\vh}_T)-L_T(\vh)
	+L_T(\vh)-\overbar L_T(\vh)\\
	&\leq Z_T.
\end{align*}
Thus it remains to control $Z_T$. The proof has three steps: derive a Lipschitz bound using
\eqref{eq:real-gradient-definition}, establish bounded differences, and control
the mean by symmetrization and chaining.

\paragraph{Real-gradient calculation and Lipschitz continuity}
By \eqref{eq:sample-gradient} and Assumption~\ref{assump:bounded-design},
\begin{align*}
	\bm0\preceq\mA_{t,i}\preceq\mI_d,
	\quad
	\bm0\preceq\mC_t(\vx)\preceq\mI_d,
\end{align*}
so
\(
\opnorm{\mC_t(\vx)-\mA_{t,i}}\leq1
\)
and
\begin{align*}
	\|\nabla\ell_t(i;\vx)\|_2
	\leq
	\frac{2\|\vx\|_2}{\tau}
	\leq
	\frac{2R}{\tau},
	\quad
	\|\vx\|_2\leq R.
\end{align*}
For $\vx,\vx'$ in the radius-$R$ parameter ball, define
$\vx_s=\vx'+s(\vx-\vx')$ for $s\in[0,1]$.  By convexity,
$\|\vx_s\|_2\leq R$.  The fundamental theorem of calculus and
\eqref{eq:real-gradient-definition} yield
\begin{align}
	|\ell_t(i;\vx)-\ell_t(i;\vx')|
	\leq
	\int_0^1
	\|\nabla\ell_t(i;\vx_s)\|_2\|\vx-\vx'\|_2\,ds
	\leq
	\frac{2R}{\tau}\|\vx-\vx'\|_2.
	\label{eq:global-lipschitz}
\end{align}

\paragraph{Concentration around the mean}
Write
\begin{align*}
	s_{t,k}(\vx):=\frac{|\va_{t,k}^{\tranH}\vx|^2}{\tau}.
\end{align*}
For any $\vz$ such that $\|\vz\|_2\leq R$, one has
\(
	0 \leq s_{t,k}(\vz) \leq \frac{R^2}{\tau},
\)
which implies that 
\begin{align*}
	\left| s_{t,i}(\vz) - s_{t,i'}(\vz )\right| \leq \frac{R^2}{\tau}, \forall i,i'\in [N]
\end{align*}
Since
\(
	\ell_t(i;\vx)
	=
	\log\sum_{j=1}^N\exp(s_{t,j}(\vx))-s_{t,i}(\vx),
\)
replacing an observation $i$ by $i'$ changes a centered loss by
\begin{align*}
	\left|
	\bigl(\ell_t(i;\vx)-\ell_t(i;\vh)\bigr)
	-
	\bigl(\ell_t(i';\vx)-\ell_t(i';\vh)\bigr)
	\right|&=
	\left|
	\bigl(s_{t,i'}(\vx)-s_{t,i}(\vx)\bigr)
	-
	\bigl(s_{t,i'}(\vh)-s_{t,i}(\vh)\bigr)
	\right|\\
	&\leq\frac{2R^2}{\tau}.
\end{align*}
Thus the bounded-difference constant of $Z_T$ in its $t$-th argument is at
most $2R^2/(\tau T)$.  Then McDiarmid's inequality yields, with probability at least
$1-\gamma$,
\begin{align}
	Z_T
	\leq
	\E{Z_T}
	+C\frac{R^2}{\tau\sqrt T}\sqrt{\log(1/\gamma)}.
	\label{eq:global-mcdiarmid}
\end{align}

\paragraph{Mean deviation}
Let $\sigma_1,\ldots,\sigma_T$ be independent Rademacher variables,
independent of the observations.  The usual ghost-sample argument, followed by
the triangle inequality and exchangeability, gives
\begin{align*}
	\E{Z_T}
	\leq
	\frac2T\E{
	\sup_{\|\vx\|_2\leq R}
	\left|
	\sum_{t=1}^T\sigma_t
	\bigl(\ell_t(I_t;\vx)-\ell_t(I_t;\vh)\bigr)
	\right|}.
\end{align*}
Define 
\begin{align*}
	X_{\vx} = \sum_{t=1}^T\sigma_t
	\bigl(\ell_t(I_t;\vx)-\ell_t(I_t;\vh)\bigr).
\end{align*}
Conditioned on $I_1,\ldots,I_T$, it can be seen that
\begin{align*}
	\E[\sigma]{
	\exp\!\left(\lambda(X_{\vx}-X_{\vx'})\right)}
	&\leq
	\exp\!\left(
	\frac{\lambda^2}{2}
	\sum_{t=1}^T
	\bigl(\ell_t(I_t;\vx)-\ell_t(I_t;\vx')\bigr)^2
	\right)\leq
	\exp\!\left(
	\frac{\lambda^2}{2}
	\frac{4TR^2}{\tau^2}\|\vx-\vx'\|_2^2
	\right),
\end{align*}
where we have used the elementary bound
$\cosh(u)\leq e^{u^2/2}$ and \eqref{eq:global-lipschitz}. Therefore, for any $\|\vx\|_2, \|\vx'\|_2\leq R$, $X_{\vx} - X_{\vx'}$ is $\nu^2$-sub-Gaussian with 
\begin{align*}
	\nu^2 = \frac{4TR^2}{\tau^2}\|\vx-\vx'\|_2^2.
\end{align*}
The complex ball is isometric to the Euclidean ball in $\R^{2d}$, whose
covering numbers satisfy
\begin{align*}
	\log\mathcal N
	\left(
	\{\vx\in\C^d:\|\vx\|_2\leq R\},
	\|\cdot\|_2,u
	\right)
	\leq
	2d\log\!\left(1+\frac{2R}{u}\right).
\end{align*}
Noting that $X_{\vh}=0$, by Dudley's entropy integral, one has
\begin{align*}
	\E[\sigma]{\sup_{\|\vx\|_2\leq R}|X_{\vx}|}
	&\leq
	C\frac{2R\sqrt T}{\tau}
	\int_0^{2R}
	\sqrt{2d\log\!\left(1+\frac{2R}{u}\right)}\,du\leq
	C\frac{R^2\sqrt{2dT}}{\tau}.
\end{align*}
The last integral is finite after the substitution $u=2Rv$.  Therefore
\begin{align}
	\E{Z_T}
	\leq
	C\frac{R^2\sqrt{2d}}{\tau\sqrt T}.
	\label{eq:global-mean}
\end{align}
Combining \eqref{eq:global-mcdiarmid} and \eqref{eq:global-mean} proves the
claim.

%%%%%%%%%%%%%%%
\subsection{Proof of Theorem~\ref{thm:risk-to-distance}}
	Theorem~\ref{thm:global} gives, for some absolute constant
	$C_{\mathrm g}>0$,
	\begin{align*}
		\overbar L_T(\widehat{\vh}_T)-\overbar L_T(\vh) &\leq 	C_{\mathrm g}\frac{R^2}{\tau}	\sqrt{\frac{d+\log(1/\eta)}{T}}\leq \frac{C_{\mathrm g}}{\sqrt C}
		\frac{\kappa_F^-\|\vh\|_2^2\rho^2}{\tau^2},
	\end{align*}
	with probability at least $1-\eta$, where the second inequality follows from the sample-size condition  \eqref{eq:entry-sample-size}. Since $C_{\mathrm g}$ is an absolute constant, we may choose the absolute
	constant $C$ in \eqref{eq:entry-sample-size} so that
	$C_{\mathrm g}/\sqrt C<1/8$; for example, it suffices to take
	$C>64C_{\mathrm g}^2$.  It then follows that
	\begin{align*}
		\overbar L_T(\widehat{\vh}_T)-\overbar L_T(\vh)
		<
		\frac{\kappa_F^-\|\vh\|_2^2\rho^2}{8\tau^2}.
	\end{align*}
	On this event, if
	$\distT(\widehat{\vh}_T,\vh)\geq\rho$, then
	Lemma~\ref{lem:fisher-frame-consequences} and
	Lemma~\ref{lem:rank-one-quotient} imply
	\begin{align*}
		\overbar L_T(\widehat{\vh}_T)-\overbar L_T(\vh)
		&\geq
		\frac{\kappa_F^-}{4\tau^2}
		\fronorm{
			\widehat{\vh}_T\widehat{\vh}_T^{\tranH}
			-\vh\vh^{\tranH}}^2\geq
		\frac{\kappa_F^-\|\vh\|_2^2}{8\tau^2}
		\distT^2(\widehat{\vh}_T,\vh)\geq
		\frac{\kappa_F^-\|\vh\|_2^2\rho^2}{8\tau^2}.
	\end{align*}
	This contradicts the strict upper bound above. Hence
	$\distT(\widehat{\vh}_T,\vh)<\rho$.

%%%%%%%%%%%%%%%%%
\subsection{Proof of Theorem~\ref{thm:local}}
	The proof has two logically separate stages: the global result places the
	MLE in the local tube, and the local likelihood expansion determines its
	statistical accuracy.
	
	\paragraph{Entry and alignment}
	Define the entry event
	\begin{align*}
		\mathcal E_{\mathrm{ent}}
		:=
		\left\{
		\distT(\widehat{\vh}_T,\vh)<\rho
		\right\}.
	\end{align*}
	The first term in \eqref{eq:local-sample-size} permits
	Theorem~\ref{thm:risk-to-distance} to be applied with
	$\eta=\gamma/3$, and hence
	\begin{align*}
		\Pr{\mathcal E_{\mathrm{ent}}}\geq1-\gamma/3.
	\end{align*}
	Define the curvature event
	\begin{align*}
		\mathcal E_{\mathrm{rsc}}
		:=
		\Bigl\{\,
		\innerR{\nabla^2L_T(\vx)[\vz]}{\vz}
		\geq
		\frac34\innerR{\mathcal I_{\vh}\vz}{\vz}\text{for every }\vx\in\mathcal N_\rho(\vh)
		\text{ and }\vz\in\mathcal H_{\vh}
		\Bigr\}.
	\end{align*}
	Lemma~\ref{lem:horizontal-fisher-frame} gives the Fisher eigenvalue bounds
	\eqref{eq:fisher-eigen-frame}.  Together with
	\eqref{eq:local-radius-condition}, these bounds imply the radius condition
	in Lemma~\ref{lem:local-rsc}; together with the second and third terms in
	\eqref{eq:local-sample-size}, they also imply its sample-size condition.
	Applying that lemma with $\delta=\gamma/3$ gives
	\begin{align*}
		\Pr{\mathcal E_{\mathrm{rsc}}}\geq1-\gamma/3.
	\end{align*}
	Define the score event
	\begin{align*}
		\mathcal E_{\mathrm{score}}
		:=
		\left\{
		\left\|
\bm I_{\bm h}^{-1/2}
\nabla L_T(\bm h)
\right\|_2^2
		\leq
		C\frac{d+\log(6/\gamma)}{T}
		\right\},
	\end{align*}
	where $C$ is the absolute constant from
	Lemma~\ref{lem:score-concentration}.  The last term in
	\eqref{eq:local-sample-size} and the lower Fisher eigenvalue bound allow
	that lemma to be applied with $\delta=\gamma/3$, yielding
	\begin{align*}
		\Pr{\mathcal E_{\mathrm{score}}}\geq1-\gamma/3.
	\end{align*}
	Finally, set
	\begin{align}
		\mathcal E
		:=
		\mathcal E_{\mathrm{ent}}
		\cap\mathcal E_{\mathrm{rsc}}
		\cap\mathcal E_{\mathrm{score}}.
		\label{eq:local-good-event}
	\end{align}
	A union bound therefore gives $\Pr{\mathcal E}\geq1-\gamma$.  Work on
	$\mathcal E$.  The alignment step requires no additional random event:
	for every realization in $\mathcal E_{\mathrm{ent}}$, let
	$\widehat{\vh}_T^{\mathrm a}$ be the optimally phase-aligned representative
	from Lemma~\ref{lem:phase-procrustes-alignment}, and set
	\begin{align*}
		\widehat\vz
		:=
		\widehat{\vh}_T^{\mathrm a}-\vh.
	\end{align*}
	Phase invariance makes $\widehat{\vh}_T^{\mathrm a}$ another empirical
	minimizer.  Moreover,
	\begin{align*}
		\widehat\vz\in\mathcal H_{\vh},
		\qquad
		\|\widehat\vz\|_2
		=
		\distT(\widehat{\vh}_T,\vh)
		<\rho,
	\end{align*}
	and every point $\vh+s\widehat\vz$, $0\leq s\leq1$, lies in
	$\mathcal N_\rho(\vh)$.
	
	\paragraph{Basic inequality in Fisher geometry}Since $\vh+\widehat{\vz}$ is an empirical minimizer, one has
	\begin{align*}
		0 &\geq L_T(\vh+\widehat\vz)-L_T(\vh)=
		\innerR{\nabla L_T(\vh)}{\widehat\vz}
		+
		\int_0^1(1-s)
		\innerR{
			\nabla^2L_T(\vh+s\widehat\vz)[\widehat\vz]
		}{\widehat\vz}\,ds\\
		&\geq \innerR{\nabla L_T(\vh)}{\widehat\vz} + \frac38 \|\widehat\vz\|_{\mathcal I_{\vh}}^2,
	\end{align*}
	where the second line follows from the Taylor's formula with integral remainder, and the last line is due to Lemma~\ref{lem:local-rsc}. Therefore,
	\begin{align*}
\frac38\|\widehat\vz\|_{\mathcal I_{\vh}}^2
		&\leq
		|\innerR{\nabla L_T(\vh)}{\widehat\vz}|\leq
		 \|\widehat\vz\|_{\mathcal I_{\vh}} \cdot \left\|
\bm I_{\bm h}^{-1/2}
\nabla L_T(\bm h)
\right\|_2,
	\end{align*}
    The lase line follows by normalizing
	$\mathcal I_{\vh}^{1/2}\widehat\vz\in\mathcal H_{\vh}$ when
	$\widehat\vz\ne0$, whose Euclidean norm is
	$\|\widehat\vz\|_{\mathcal I_{\vh}}$.
	Thus, including the trivial case $\widehat\vz=0$,
	\begin{align}
		\|\widehat\vz\|_{\mathcal I_{\vh}}^2
		\leq
		\frac{64}{9}\left\|
\bm I_{\bm h}^{-1/2}
\nabla L_T(\bm h)
\right\|_2^2
		\leq
		C\frac{d+\log(1/\gamma)}{T},
		\label{eq:local-proof-fisher}
	\end{align}
	where the last step uses Lemma~\ref{lem:score-concentration}.  Since
	$\mathcal I_{\vh}\succeq\lambda_-\mathsf I$ on
	$\mathcal H_{\vh}$,
	\begin{align*}
		\distT^2(\widehat{\vh}_T,\vh)
		&=
		\|\widehat\vz\|_2^2	\leq	\lambda_-^{-1}	\|\widehat\vz\|_{\mathcal I_{\vh}}^2\leq	\frac{\tau^2}{\kappa_F^-\|\vh\|_2^2}	\|\widehat\vz\|_{\mathcal I_{\vh}}^2\leq C\frac{\tau^2\{d+\log(1/\gamma)\}}
		{T\kappa_F^-\|\vh\|_2^2},
	\end{align*}
	which proves
	\eqref{eq:local-parameter-frame}.
	
	\paragraph{Population excess risk}
	Notice that $\nabla\overbar L_T(\vh)=0$.  The population
	Hessian Lipschitz bound \eqref{eq:population-hessian-lipschitz} and
	\eqref{eq:local-radius-condition} imply, for $0\leq s\leq1$,
	\begin{align*}
		\innerR{\nabla^2\overbar L_T(\vh+s\widehat\vz)
			[\widehat\vz]}{\widehat\vz}
		&\leq
		\|\widehat\vz\|_{\mathcal I_{\vh}}^2
		+L_H\rho\|\widehat\vz\|_2^2\leq
		\frac98\|\widehat\vz\|_{\mathcal I_{\vh}}^2.
	\end{align*}
	A second Taylor expansion therefore gives
	\begin{align*}
		\overbar L_T(\widehat{\vh}_T)-\overbar L_T(\vh)
		&=
		\int_0^1(1-s)
		\innerR{\nabla^2\overbar L_T(\vh+s\widehat\vz)
			[\widehat\vz]}{\widehat\vz}\,ds\leq
		\frac9{16}\|\widehat\vz\|_{\mathcal I_{\vh}}^2\leq C\frac{d+\log(1/\gamma)}{T},
	\end{align*}
	which proves \eqref{eq:local-excess-fisher}.

%%%%%%%%%%%%%%%%%%%
\subsection{Proof of Theorem~\ref{thm:minimax-lower}}
 The proof follows from a standard procedure in statistical learning, see for example~\cite{yu1997assouad,wainwright2019high}.
 It first reduces estimation to testing for finite packing and then achieves a lower bound based on the mutual information.  The concrete packing points are constructed in a local complex
 cross-section via the Varshamov--Gilbert lemma \cite[Lemma 2.9]{tsybakov2009introduction}, and proper adjustments of the parameters lead to the desired bound.
 
 \paragraph{Lower bound via mutual information}
 Let
 \(
 \{\vh_\omega\}_{\omega\in\Omega}
 \subseteq
 \Theta(\vh_0,\rho)
 \)
 be a finite $2\Delta$-packing under $\distT$, so that
 \[
 \distT(\vh_\omega,\vh_{\omega'})
 \geq2\Delta,
 \quad
 \omega\neq\omega'.
 \]
 Let $W$ be a random variable uniformly distributed on $\Omega$,
 \[
 \mathbb{P}(W=\omega)=\frac{1}{|\Omega|}.
 \]
 Conditioned on $W=\omega$, let $I^{(T)}$ denote the $T$ index samples from $p_t(\cdot;\vh_\omega)$, i.e.,
 \begin{align*}
 	I^{(T)}=(I_1,\ldots,I_T)~|~(W=\omega)
 	\sim P^{(T)}_{\vh_\omega}:=\bigotimes_{t=1}^T p_t(\cdot;\vh_\omega).
 \end{align*}
 By a standard argument and Fano's inequality, one has
 \[
 \mathcal{R}_T(\vh_0,\rho)\geq \Delta^2\left(1-\frac{I(W;I^{(T)})+\log 2}{\log |\Omega|}\right).
 \]
 Moreover, a direct calculation yields
 \[
 I(W;I^{(T)})
 =\frac{1}{|\Omega|}\sum_{\omega\in\Omega}
 \KL(P_{\vh_\omega}^{(T)}\|\overbar P^{(T)})
 \leq \frac{1}{|\Omega|}\sum_{\omega\in\Omega}
 \KL(P_{\vh_\omega}^{(T)}\|P_{\vh_0}^{(T)}),
 \]
 where
 \(
 \overbar P^{(T)}=|\Omega|^{-1}\sum_{\omega\in\Omega}P_{\vh_\omega}^{(T)}
 \).
 Therefore,
 \[
 \mathcal{R}_T(\vh_0,\rho)
 \geq \Delta^2\left(
 1-\frac{|\Omega|^{-1}\sum_{\omega\in\Omega}
 	\KL(P_{\vh_\omega}^{(T)}\|P_{\vh_0}^{(T)})+\log 2}
 {\log |\Omega|}
 \right).
 \]
 
 \paragraph{Construction of $\Omega$}
 Recall that $r_0=\|\vh_0\|_2$ and $r_+=r_0+\rho$. Let $m=2d-1$ be the dimension of
 \begin{align*}
 	\mathcal H_{\vh_0}
 	:=
 	\left\{
 	\vz\in\C^d:
 	\innerR{\mathrm i\vh_0}{\vz}=0
 	\right\},
 \end{align*}
 and let $\vb_1,\ldots,\vb_m$ be an orthonormal basis of $\mathcal H_{\vh_0}$ under the real inner product $\innerR{\cdot}{\cdot}$. By the Varshamov--Gilbert lemma, there exists a subset
 \(
 \Omega\subset\{0,1\}^m
 \)
 such that $\log|\Omega|\geq m\log 2/8$ and the Hamming distance satisfies
 \[
 d_{\mathrm H}(\omega,\omega')\geq \frac{m}{8},
 \qquad
 \omega\neq\omega'.
 \]
 Define
 \[
 \vz_\omega=a\sum_{j=1}^m\omega_j\vb_j,
 \qquad
 \vh_\omega=\vh_0+\vz_\omega.
 \]
 It is easy to verify that
 \begin{align*}
 	\|\vz_\omega-\vz_{\omega'}\|_2^2
 	=a^2d_{\mathrm H}(\omega,\omega').
 \end{align*}
 Similarly,
 \[
 \|\vz_\omega\|_2^2=a^2\|\omega\|_0\leq a^2m.
 \]
 Therefore, $\vh_\omega\in\Theta(\vh_0,\rho)$ provided $a^2m\leq\rho^2$.
 
 Let
 \[
 \vh=\vh_0+\vz,
 \qquad
 \vh'=\vh_0+\vz',
 \qquad
 \vz,\vz'\in\mathcal H_{\vh_0},
 \qquad
 \|\vz\|_2,\|\vz'\|_2\leq\rho.
 \]
 Define
 \(
 \bm\delta=\vz-\vz'
 \)
 and write
 \(
 \vh^{\tranH}\vh'=x+\mathrm i y
 \).
 Then
 \begin{align*}
 	\distT^2(\vh,\vh')
 	&=\|\vh\|_2^2+\|\vh'\|_2^2-2|\vh^{\tranH}\vh'|\\
 	&=\|\vh-\vh'\|_2^2
 	-2\left(|\vh^{\tranH}\vh'|-\operatorname{Re}(\vh^{\tranH}\vh')\right)\\
 	&=\|\bm\delta\|_2^2
 	-\frac{2y^2}{\sqrt{x^2+y^2}+x}.
 \end{align*}
 Note that
 \begin{align*}
 	x
 	=\operatorname{Re}(\vh^{\tranH}\vh')
 	=\vh_0^{\tranH}\vh_0+\vh_0^{\tranH}\vz'
 	+\vz^{\tranH}\vh_0+\operatorname{Re}(\vz^{\tranH}\vz')
 	\geq r_0^2-2r_0\rho-\rho^2
 	\geq\frac{7r_0^2}{16},
 \end{align*}
 where the last inequality follows from $\rho\leq r_0/4$. Consequently,
 \[
 \distT^2(\vh,\vh')
 \geq \|\bm\delta\|_2^2-\frac{y^2}{x}.
 \]
 In addition, since
 \begin{align*}
 	y
 	&=\operatorname{Im}(\vh^{\tranH}\vh')
 	=\operatorname{Im}(\vz^{\tranH}\vz')
 	=\operatorname{Im}((\vz-\vz')^{\tranH}\vz'),
 \end{align*}
 there holds
 \(
 |y|\leq\|\bm\delta\|_2\rho
 \).
 Therefore,
 \[
 \distT^2(\vh,\vh')
 \geq\left(1-\frac{16\rho^2}{7r_0^2}\right)
 \|\bm\delta\|_2^2
 \geq\frac67\|\bm\delta\|_2^2.
 \]
 It follows that
 \begin{align*}
 	\distT^2(\vh_\omega,\vh_{\omega'})
 	&\geq\frac67a^2d_{\mathrm H}(\omega,\omega')
 	\geq\frac67\frac{a^2m}{8}
 	=4\Delta^2,
 \end{align*}
 where
 \[
 \Delta^2=\frac67\frac{a^2m}{32}.
 \]
 Note that this requires $a^2m\leq\rho^2$ and $\rho\leq r_0/4$.

 	\paragraph{KL divergence bound}
 	Due to conditional independence,
 	\begin{align*}
 		\KL(P_{\vh_\omega}^{(T)}\|P_{\vh_0}^{(T)})
 		=\sum_{t=1}^T
 		\KL(p_t(\cdot;\vh_\omega)\|p_t(\cdot;\vh_0)).
 	\end{align*}
 	For each $\omega\in\Omega$, set
 	\begin{align*}
 		\mX_\omega
 		&:=
 		\vh_\omega\vh_\omega^{\tranH},
 		\quad
 		\mH_0
 		:=
 		\vh_0\vh_0^{\tranH},\quad
 		\bm\Delta_\omega
 		:=
 		\mX_\omega-\mH_0.
 	\end{align*}
 	Because $\|\vh_\omega\|_2\leq r_+\leq R$ and
 	$\|\vh_0\|_2=r_0\leq R$, for every $s\in[0,1]$,
 	\begin{align*}
 		\mX_\omega-s\bm\Delta_\omega
 		=
 		(1-s)\vh_\omega\vh_\omega^{\tranH}
 		+s\vh_0\vh_0^{\tranH}
 	\end{align*}
 	is a convex combination of rank-one lifts generated by vectors with norms
 	at most $R$, while
 	$\bm\Delta_\omega=\vh_\omega\vh_\omega^{\tranH}
 	-\vh_0\vh_0^{\tranH}$ is the difference of two such lifts.  Thus the
 	hypotheses of the upper inequality in Assumption~\ref{assump:fisher-frame}
 	hold along the entire path.  Moreover,
 	$p_t(\cdot;\vh_\omega)=q_t(\cdot;\mX_\omega)$ and
 	$p_t(\cdot;\vh_0)=q_t(\cdot;\mH_0)$.  For one round,
 	\begin{align*}
 		\KL\bigl(p_t(\cdot;\vh_\omega)\,\|\, p_t(\cdot;\vh_0)\bigr)	&=
 		f_t(\mH_0)-f_t(\mX_\omega)
 		-\innerR{\nabla f_t(\mX_\omega)}
 		{\mH_0-\mX_\omega}\\
 		&=
 		\int_0^1(1-s)
 		\innerR{\nabla^2f_t(\mX_\omega-s\bm\Delta_\omega)
 			[\bm\Delta_\omega]}{\bm\Delta_\omega}\,ds.
 	\end{align*}
 	Summing over the independent rounds and applying
 	\eqref{eq:lifted-hessian} and the upper inequality in
 	\eqref{eq:fisher-frame} along this path yields
 	\begin{align*}
 		\KL(P_{\vh_\omega}^{(T)}\,\|\,P_{\vh_0}^{(T)})
 		&\leq
 		\frac{T\kappa_F^+}{4\tau^2}
 		\fronorm{\bm\Delta_\omega}^2.
 	\end{align*}
 	Finally,
 	\begin{align*}
 		\fronorm{\bm\Delta_\omega}
 		&\leq
 		(\|\vh_\omega\|_2+\|\vh_0\|_2)
 		\|\vh_\omega-\vh_0\|_2\leq
 		2r_+\|\vz_\omega\|_2,
 	\end{align*}
 	and hence
 	\begin{align*}
 		\KL(P_{\vh_\omega}^{(T)}\,\|\,P_{\vh_0}^{(T)})
 		\leq
 		\frac{T\kappa_F^+r_+^2}{\tau^2}
 		\|\vz_\omega\|_2^2.
 	\end{align*}
 	
 	\paragraph{Put it all together}
 	The preceding KL bound and $\|\vz_\omega\|_2^2\leq a^2m$ give
 	\begin{align*}
 		\mathcal R_T(\vh_0,\rho)
 		&\geq
 		\Delta^2\left(
 		1-\frac{|\Omega|^{-1}\sum_{\omega\in\Omega}
 			\KL(P_{\vh_\omega}^{(T)}\|P_{\vh_0}^{(T)})+\log2}
 		{\log|\Omega|}
 		\right)\geq
 		\Delta^2\left(
 		1-\frac{8T\kappa_F^+r_+^2a^2}{\tau^2\log2}
 		-\frac8m
 		\right).
 	\end{align*}
 	Set
 	\begin{align*}
 		a^2
 		:=
 		c_0\min\left\{
 		\frac{\rho^2}{m},
 		\frac{\tau^2}{T\kappa_F^+r_+^2}
 		\right\},
 		\qquad
 		c_0:=\frac{\log2}{128}.
 	\end{align*}
 	Then $a^2m\leq\rho^2$. Moreover, $m=2d-1\geq17$ implies
 	\begin{align*}
 		1-\frac{8T\kappa_F^+r_+^2a^2}{\tau^2\log2}
 		-\frac8m
 		\geq\frac7{16}.
 	\end{align*}
 	Consequently,
 	\begin{align*}
 		\mathcal R_T(\vh_0,\rho)
 		&\geq\frac7{16}\Delta^2=\frac{3\log2}{32768}
 		\min\left\{
 		\rho^2,
 		\frac{\tau^2m}{T\kappa_F^+r_+^2}
 		\right\}.
 	\end{align*}
 	Since $m=2d-1\geq d$, it follows that
 	\begin{align*}
 		\mathcal R_T(\vh_0,\rho)
 		\geq
 		c
 		\min\left\{
 		\rho^2,
 		\frac{\tau^2d}{T\kappa_F^+r_+^2}
 		\right\},
 	\end{align*}
 	where $c>0$ is an absolute constant.  This completes the proof.

%%%%%%%%%%%%%%%%%%%%%%%
\section{Proofs of Technical Lemmas}
\label{sec:proofs-lemmas}
\subsection{Proof of Lemma~\ref{lem:complex-secant-stiefel}}
	Let
	\begin{align*}
		\mathbb H_0^d
		:=
		\left\{
		\mM\in\C^{d\times d}:
		\mM=\mM^{\tranH},\ \operatorname{tr}(\mM)=0
		\right\}.
	\end{align*}
	Equipped with the real inner product
	$\innerR{\mM}{\mC}=\operatorname{Re}\operatorname{tr}(\mM^{\tranH}\mC)$,
	this is a real Hilbert space of dimension
	\begin{align*}
		m:=\dim_{\R}(\mathbb H_0^d)=d^2-1.
	\end{align*}
	Because the codewords have unit norm and
	$\mQ_t^{\tranH}\mQ_t=\mI_p$, one has $\|\va_{t,i}\|_2=1$.  Consequently,
	each difference
	\begin{align*}
		\mB_{t,ij}:=\mA_{t,i}-\mA_{t,j}
	\end{align*}
	is Hermitian and traceless.
	
	For each $t$, define the self-adjoint positive semidefinite operator
	$\mathcal V_t:\mathbb H_0^d\to\mathbb H_0^d$ by
	\begin{align*}
		\mathcal V_t(\mM)
		:=
		\sum_{i=1}^N\sum_{j=1}^N
		\innerR{\mB_{t,ij}}{\mM}\mB_{t,ij}.
	\end{align*}
	Its quadratic form is
	\begin{align}
		\innerR{\mathcal V_t(\mM)}{\mM}
		=
		\sum_{i=1}^N\sum_{j=1}^N
		\innerR{\mB_{t,ij}}{\mM}^2.
		\label{eq:complex-operator-quadratic-form}
	\end{align}
	
	We first compute the mean of $\mathcal V_t$.  Set
	\begin{align*}
		\mJ
		:=
		\begin{bmatrix}
			\mI_p\\
			\bm 0
		\end{bmatrix}
		\in\C^{d\times p}.
	\end{align*}
	A Haar-distributed matrix on $\operatorname{St}(d,p)$ has the same law
	as $\mU\mJ$, where $\mU$ is Haar-distributed on the unitary group
	$\mathrm U(d)$.  Hence, with
	\begin{align*}
		\mC_{ij}
		:=
		\mJ(\vv_i\vv_i^{\tranH}-\vv_j\vv_j^{\tranH})\mJ^{\tranH}
		\in\mathbb H_0^d,
	\end{align*}
	one has $\mB_{t,ij}\overset{\mathrm d}=\mU\mC_{ij}\mU^{\tranH}$ and
	\begin{align*}
		\fronorm{\mC_{ij}}
		=
		\fronorm{\vv_i\vv_i^{\tranH}-\vv_j\vv_j^{\tranH}}.
	\end{align*}
	
	For any fixed $\mC,\mM\in\mathbb H_0^d$, Haar invariance gives the
	isotropy identity
	\begin{align}
		\E[\mU]{
			\innerR{\mU\mC\mU^{\tranH}}{\mM}^2
		}
		=
		\frac{\fronorm{\mC}^2}{d^2-1}\fronorm{\mM}^2.
		\label{eq:complex-haar-isotropy}
	\end{align}
	For completeness, diagonalize $\mC$ and absorb its eigenvectors into the
	Haar matrix.  Thus, without loss of generality,
	\begin{align*}
		\mC=\operatorname{diag}(\lambda_1,\ldots,\lambda_d),
		\qquad
		\sum_{r=1}^d\lambda_r=0.
	\end{align*}
	Writing $\mU=[\vu_1,\ldots,\vu_d]$ and
	$X_r:=\vu_r^{\tranH}\mM\vu_r$, unitary invariance yields constants $a$ and
	$b$ such that
	\begin{align*}
		\E{X_r^2}=a,
		\qquad
		\E{X_rX_s}=b,
		\quad r\neq s.
	\end{align*}
	Since $\sum_rX_r=\operatorname{tr}(\mM)=0$, one has
	$b=-a/(d-1)$.  If $\vu$ is uniformly distributed on the complex unit sphere,
	its fourth moment satisfies
	\begin{align*}
		\E{
			\overline{u}_au_b\overline{u}_cu_e
		}
		=
		\frac{\delta_{ab}\delta_{ce}+\delta_{ae}\delta_{cb}}
		{d(d+1)}.
	\end{align*}
	This identity follows by writing
	$\vu=\vg/\|\vg\|_2$ for a standard  complex Gaussian vector
	$\vg$.  It implies
	\begin{align*}
		a
		=
		\E{(\vu^{\tranH}\mM\vu)^2}
		=
		\frac{(\operatorname{tr}\mM)^2+\operatorname{tr}(\mM^2)}{d(d+1)}
		=
		\frac{\fronorm{\mM}^2}{d(d+1)}.
	\end{align*}
	Using
	$\sum_{r\neq s}\lambda_r\lambda_s=-\sum_r\lambda_r^2
	=-\fronorm{\mC}^2$, we obtain
	\begin{align*}
		\E[\mU]{
			\innerR{\mU\mC\mU^{\tranH}}{\mM}^2
		}
		&=
		(a-b)\fronorm{\mC}^2=
		\frac{\fronorm{\mC}^2\fronorm{\mM}^2}{(d-1)(d+1)},
	\end{align*}
	which proves \eqref{eq:complex-haar-isotropy}.
	
	It now follows from \eqref{eq:complex-operator-quadratic-form} and
	\eqref{eq:complex-haar-isotropy} that
	\begin{align}
		\E{\mathcal V_t}
		=
		\mu_V\mathcal I_{\mathbb H_0^d},
		\qquad
		\mu_V
		:=
		\frac1m\sum_{i=1}^N\sum_{j=1}^N
		\fronorm{\vv_i\vv_i^{\tranH}-\vv_j\vv_j^{\tranH}}^2.
		\label{eq:complex-operator-mean}
	\end{align}

		By $m=d^2-1$ and the definition \eqref{eq:codebook-spread},
		\begin{align}
			\mu_V
			=
			\frac{N^2s_{\mathcal V}}{d^2-1}.
			\label{eq:complex-mu-spread}
		\end{align}

	Moreover, for every $\mM\in\mathbb H_0^d$,
	\begin{align*}
		\innerR{\mathcal V_t(\mM)}{\mM}
		&\leq
		\left(
		\sum_{i=1}^N\sum_{j=1}^N\fronorm{\mB_{t,ij}}^2
		\right)\fronorm{\mM}^2= m\mu_V\fronorm{\mM}^2.
	\end{align*}
	Thus
	\begin{align*}
		0\preceq\mathcal V_t
		\preceq
		m\mu_V\mathcal I_{\mathbb H_0^d}.
	\end{align*}
	
	The operators $\mathcal V_1,\ldots,\mathcal V_T$ are independent.
	The matrix Chernoff inequality, applied on the $m$-dimensional real Hilbert
	space $\mathbb H_0^d$, therefore gives
	\begin{align*}
		\Pr{
			\lambda_{\min}\left(\frac1T\sum_{t=1}^T\mathcal V_t\right)
			\leq(1-\delta)\mu_V
		}\leq
		m\left[
		\frac{e^{-\delta}}{(1-\delta)^{1-\delta}}
		\right]^{T/m}
		\leq
		m\exp\left(-\frac{\delta^2T}{2m}\right).
	\end{align*}
	Consequently, outside an event of probability at most
	$(d^2-1)\exp(-\delta^2T/(2(d^2-1)))$, one has
	\begin{align}
		\frac1T\sum_{t=1}^T\sum_{i=1}^N\sum_{j=1}^N
		\innerR{\mB_{t,ij}}{\mM}^2
		\geq
		(1-\delta)\mu_V\fronorm{\mM}^2,
		\qquad
		\mM\in\mathbb H_0^d.
		\label{eq:complex-traceless-frame-bound}
	\end{align}
		The upper-tail matrix Chernoff inequality also gives
		\begin{align*}
			\Pr{
				\lambda_{\max}\left(\frac1T\sum_{t=1}^T\mathcal V_t\right)
				\geq(1+\delta)\mu_V
			}
			\leq
			m\exp\left(-\frac{\delta^2T}{3m}\right).
		\end{align*}
		Intersecting the lower and upper tail events, with probability at least
		\begin{align*}
			1-2(d^2-1)
			\exp\!\left(-\frac{\delta^2T}{3(d^2-1)}\right),
		\end{align*}
		the lower bound in \eqref{eq:complex-traceless-frame-bound} holds and,
		simultaneously,
		\begin{align}
			\frac1T\sum_{t=1}^T\sum_{i=1}^N\sum_{j=1}^N
			\innerR{\mB_{t,ij}}{\mM}^2
			\leq
			(1+\delta)\mu_V\fronorm{\mM}^2,
			\qquad
			\mM\in\mathbb H_0^d.
			\label{eq:complex-traceless-frame-upper}
		\end{align}

	The coherence assumption supplies an explicit lower bound for $\mu_V$.
	Indeed, for $i\neq j$,
	\begin{align*}
		\fronorm{\vv_i\vv_i^{\tranH}-\vv_j\vv_j^{\tranH}}^2
		=
		2\left(1-|\vv_i^{\tranH}\vv_j|^2\right)
		\geq
		2(1-\mu^2).
	\end{align*}
	There are $N(N-1)$ ordered pairs with $i\neq j$, and hence
	\begin{align}
		\mu_V
		\geq
		\frac{2N(N-1)(1-\mu^2)}{d^2-1}.
		\label{eq:complex-mu-lower-bound}
	\end{align}
		The same identity also gives the two-sided codebook-spread estimate in the
		lemma statement.  Indeed, the $N$ diagonal terms vanish and every
		off-diagonal term is at most $2$, so
		\begin{align*}
			2\left(1-\frac1N\right)(1-\mu^2)
			\leq s_{\mathcal V}
			\leq2\left(1-\frac1N\right)
			\leq2.
		\end{align*}

	It remains to pass from traceless Hermitian matrices to lifted secants.  Fix
	$\vx,\vh\in\C^d$ and set
	\begin{align*}
		\bm\Delta
		:=
		\vx\vx^{\tranH}-\vh\vh^{\tranH},
		\qquad
		\mathcal P_0(\bm\Delta)
		:=
		\bm\Delta-
		\frac{\operatorname{tr}(\bm\Delta)}d\mI_d.
	\end{align*}
	Since every $\mB_{t,ij}$ is traceless,
	\begin{align*}
		\innerR{\mB_{t,ij}}{\bm\Delta}
		=
		\innerR{\mB_{t,ij}}{\mathcal P_0(\bm\Delta)}.
	\end{align*}
	Furthermore,
	\begin{align*}
		\fronorm{\mathcal P_0(\bm\Delta)}^2
		=
		\fronorm{\bm\Delta}^2
		-
		\frac{(\operatorname{tr}\bm\Delta)^2}{d}.
	\end{align*}
	The Cauchy--Schwarz inequality gives
	\begin{align*}
		(\operatorname{tr}\bm\Delta)^2
		&=
		\left(\|\vx\|_2^2-\|\vh\|_2^2\right)^2\leq
		\|\vx\|_2^4+\|\vh\|_2^4-2|\vx^{\tranH}\vh|^2
		=
		\fronorm{\bm\Delta}^2.
	\end{align*}
	Therefore,
	\begin{align}
		\fronorm{\mathcal P_0(\bm\Delta)}^2
		\geq
		\left(1-\frac1d\right)\fronorm{\bm\Delta}^2.
		\label{eq:complex-secant-projection}
	\end{align}
	
	Applying \eqref{eq:complex-traceless-frame-bound} to
	$\mathcal P_0(\bm\Delta)$ and then using
	\eqref{eq:complex-mu-lower-bound}--\eqref{eq:complex-secant-projection}
	yields
	\begin{align*}
		\frac1T\sum_{t=1}^T\sum_{i=1}^N\sum_{j=1}^N
		\innerR{\mA_{t,i}-\mA_{t,j}}{\bm\Delta}^2
		&\geq
		(1-\delta)
		\frac{2N(N-1)(1-\mu^2)}{d^2-1}
		\left(1-\frac1d\right)
		\fronorm{\bm\Delta}^2\\
		&=
		(1-\delta)
		\frac{2N(N-1)(1-\mu^2)}{d(d+1)}
		\fronorm{\bm\Delta}^2.
	\end{align*}
	The event on which this argument holds does not depend on $\vx$ or $\vh$,
	so the conclusion is simultaneous over all lifted secants.
	
		It remains to establish the two-sided weighted inequalities in
		Assumption~\ref{assump:fisher-frame}.  On the simultaneous lower and
		upper tail event above, \eqref{eq:complex-traceless-frame-bound},
		\eqref{eq:complex-traceless-frame-upper}, and
		\eqref{eq:complex-secant-projection} give
		\begin{align*}
			(1-\delta)\mu_V\left(1-\frac1d\right)
			\fronorm{\bm\Delta}^2
			&\leq
			\frac1T\sum_{t=1}^T\sum_{i=1}^N\sum_{j=1}^N
			\innerR{\mB_{t,ij}}{\bm\Delta}^2\leq
			(1+\delta)\mu_V\fronorm{\bm\Delta}^2.
		\end{align*}
		Now restrict to secants generated by vectors with norms at most $R$, and
		let $\mM$ admit the
		representation specified in Assumption~\ref{assump:fisher-frame}.  Then
		$\mM$ is positive semidefinite and
		\begin{align*}
			\operatorname{tr}(\mM)
			=
			\sum_{k=1}^K\lambda_k\|\vx_k\|_2^2
			\leq R^2.
		\end{align*}
		Since
		$\mA_{t,i}=\va_{t,i}\va_{t,i}^{\tranH}$ and
		$\|\va_{t,i}\|_2=1$, we have
		\begin{align*}
			0
			\leq
			\innerR{\mA_{t,i}}{\mM}
			=
			\va_{t,i}^{\tranH}\mM\va_{t,i}
			\leq
			\|\mM\|_{\mathrm{op}}
			\leq
			\operatorname{tr}(\mM)
			\leq R^2.
		\end{align*}
		Thus every exponential appearing in the definition of $q_t(i;\mM)$ lies
		between $1$ and $e^{R^2/\tau}$.  In particular,
		\begin{align*}
			N
			\leq
			\sum_{j=1}^N
			\exp\!\left(\innerR{\mA_{t,j}}{\mM}/\tau\right)
			\leq
			Ne^{R^2/\tau}.
		\end{align*}
		Bounding the numerator and denominator separately therefore gives,
		uniformly in $t$, $i$, and $\mM$,
		\begin{align*}
			\frac{e^{-R^2/\tau}}{N}
			\leq q_t(i;\mM)
			\leq
			\frac{e^{R^2/\tau}}{N}.
		\end{align*}
		Consequently, for every $i,j$,
		\begin{align*}
			\frac{e^{-2R^2/\tau}}{N^2}
			\leq
			q_t(i;\mM)q_t(j;\mM)
			\leq
			\frac{e^{2R^2/\tau}}{N^2}.
		\end{align*}
		Because each quantity
		$\innerR{\mB_{t,ij}}{\bm\Delta}^2$ is nonnegative, the preceding
		pointwise bounds and the unweighted frame inequality imply
		\begin{align*}
			\frac{e^{-2R^2/\tau}}{N^2}(1-\delta)\mu_V
			\left(1-\frac1d\right)\fronorm{\bm\Delta}^2&\leq
			\frac1T\sum_{t=1}^T\sum_{i=1}^N\sum_{j=1}^N
			q_t(i;\mM)q_t(j;\mM)
			\innerR{\mB_{t,ij}}{\bm\Delta}^2\\
			&\leq
			\frac{e^{2R^2/\tau}}{N^2}(1+\delta)\mu_V
			\fronorm{\bm\Delta}^2.
		\end{align*}
		Using \eqref{eq:complex-mu-spread} identifies the two constants in
		this display with \eqref{eq:kappa-f-minus} and
		\eqref{eq:kappa-f-plus}, respectively.  The simultaneous lower and
		upper tail event holds for every traceless Hermitian matrix; all subsequent
		estimates are deterministic and uniform.  Hence the resulting weighted
		inequalities hold on the same event
		for every $\mM$ and $\bm\Delta$ satisfying the respective representation
		and secant conditions in Assumption~\ref{assump:fisher-frame}.

\subsection{Proof of Lemma \ref{lemma-differential}}
	We begin with the derivative of the softmax weights.  Write
	\begin{align*}
		s_{t,i}(\vx)
		:=
		\frac{\vx^{\tranH}\mA_{t,i}\vx}{\tau},
		\qquad
		Z_t(\vx)
		:=
		\sum_{j=1}^N e^{s_{t,j}(\vx)},
	\end{align*}
	so that $p_t(i;\vx)=e^{s_{t,i}(\vx)}/Z_t(\vx)$.  By
	\eqref{eq:quadratic-real-gradient}, one has
	\begin{align*}
		Ds_{t,i}(\vx)[\vz]
		=
		\frac{2}{\tau}\innerR{\mA_{t,i}\vx}{\vz}.
	\end{align*}
	Hence,
	\begin{align*}
		Dp_t(i;\vx)[\vz]
		&=
		p_t(i;\vx)
		\left(
		Ds_{t,i}(\vx)[\vz]
		-
		\sum_{j=1}^Np_t(j;\vx)Ds_{t,j}(\vx)[\vz]
		\right)=
		\frac{2p_t(i;\vx)}{\tau}
		\innerR{(\mA_{t,i}-\mC_t(\vx))\vx}{\vz}.
	\end{align*}
	
	Since $\ell_t(i;\vx)=-\log p_t(i;\vx)$, the chain rule yields
	\begin{align*}
		D\ell_t(i;\vx)[\vz]
		&=
		-\frac{Dp_t(i;\vx)[\vz]}{p_t(i;\vx)} =
		\frac{2}{\tau}
		\innerR{(\mC_t(\vx)-\mA_{t,i})\vx}{\vz},
	\end{align*}
	which implies that
	\begin{align*}
		\nabla\ell_t(i;\vx)
		=
		\frac{2}{\tau}
		(\mC_t(\vx)-\mA_{t,i})\vx.
	\end{align*}
	Averaging over the observations gives
	\begin{align*}
		\nabla L_T(\vx) = \frac{1}{T}\sum_{t=1}^{T} \nabla \ell_t(I_t; \vx)
		=
		\frac{2}{\tau T}
		\sum_{t=1}^T
		(\mC_t(\vx)-\mA_{t,I_t})\vx.
	\end{align*}
	
	We next differentiate this gradient in the direction $\vz$.  The product
	rule gives
	\begin{align*}
		D\{\mC_t(\vx)\vx\}[\vz] &=\mC_t(\vx)\vz	+(D\mC_t(\vx)[\vz])\,\vx= \mC_t(\vx)\vz	+ \sum_{i=1}^{N} Dp_t(i; \vx)[\vz]\mA_{t,i} \vx\\
		&=\mC_t(\vx)\vz	+ \sum_{i=1}^{N} \frac{2p_t(i;\vx)}{\tau}
		\innerR{(\mA_{t,i}-\mC_t(\vx))\vx}{\vz}\mA_{t,i}  \vx\\
		&=\mC_t(\vx)\vz	+ \frac{2}{\tau} \sum_{i=1}^{N} p_t(i;\vx)
		\innerR{(\mA_{t,i}-\mC_t(\vx))\vx}{\vz}\mA_{t,i}  \vx\\
		&= \mC_t(\vx)\vz	+ \frac{2}{\tau} \left(\mS_t(\vx)[\vz]	-	\innerR{\vy_t(\vx)}{\vz}\vy_t(\vx)	\right).
	\end{align*}
	Moreover, the term $\mA_{t,I_t}\vx$ has directional derivative
	$\mA_{t,I_t}\vz$.  Hence
	\begin{align*}
		\nabla^2L_T(\vx)[\vz] &= D(\nabla L_T(\vx))[\vz]=\frac{2}{\tau T} \sum_{t=1}^{T} \left(D\{ \mC_t(\vx)\vx \}[\vz] - D\{\mA_{t,I_t} \vx \}[\vz] \right)  \\
		&=
		-\frac2{\tau T}\sum_{t=1}^T\mA_{t,I_t}\vz
		+\frac2{\tau T}\sum_{t=1}^T\mC_t(\vx)\vz
		+\frac4{\tau^2T}\sum_{t=1}^T
		\left(
		\mS_t(\vx)[\vz]
		-\innerR{\vy_t(\vx)}{\vz}\vy_t(\vx)
		\right).
	\end{align*}
	
	The population expectation is taken under the true channel $\vh$.  Therefore
	\begin{align*}
		\E{\mA_{t,I_t}}
		=
		\sum_{i=1}^Np_t(i;\vh)\mA_{t,i}
		=
		\mC_t(\vh).
	\end{align*}
	The remaining terms in the empirical gradient and Hessian depend only on the
	fixed design and the evaluation point.  Taking expectations gives
	\begin{align*}
		\nabla\overbar L_T(\vx)
		&=
		\frac{2}{\tau T}
		\sum_{t=1}^T
		(\mC_t(\vx)-\mC_t(\vh))\vx,\\
		\nabla^2\overbar L_T(\vx)[\vz]
		&=
		-\frac2{\tau T}\sum_{t=1}^T\mC_t(\vh)\vz
		+\frac2{\tau T}\sum_{t=1}^T\mC_t(\vx)\vz
		+\frac4{\tau^2T}\sum_{t=1}^T
		\left(
		\mS_t(\vx)[\vz]
		-\innerR{\vy_t(\vx)}{\vz}\vy_t(\vx)
		\right).
	\end{align*}
	Given \eqref{eq:population-hessian}, it can be seen that
	\begin{align*}
		\nabla^2 \overbar{L}_T(\vh)[\vz] &=-\frac2{\tau T}\sum_{t=1}^T\mC_t(\vh)\vz
		+\frac2{\tau T}\sum_{t=1}^T\mC_t(\vh)\vz
		+\frac4{\tau^2T}\sum_{t=1}^T
		\left(
		\mS_t(\vh)[\vz]
		-\innerR{\vy_t(\vh)}{\vz}\vy_t(\vh)
		\right)\\
		&=\frac4{\tau^2T}\sum_{t=1}^T
		\left(
		\mS_t(\vh)[\vz]
		-\innerR{\vy_t(\vh)}{\vz}\vy_t(\vh)
		\right)
	\end{align*}
	Let $\alpha_{t,i}=\innerR{\mA_{t,i}\vh}{\vz}$. By the definition of $\mC_t(\vh), \mS_t(\vh)$ and $\vy_t(\vh)$, one has
	\begin{align*}
		\langle \mS_t(\vx)[\vz], \vz \rangle_{\R}	&=
		\sum_{i=1}^Np_t(i;\vx)
		\innerR{\mA_{t,i}\vx}{\vz}^2,\\
		\innerR{\vy_t(\vh)}{\vz} &= \innerR{\mC_t(\vh)\vh}{\vz} = \sum_{i=1}^{N} p_t(i; \vh) \innerR{\mA_{t,i} \vh}{\vz} =\sum_{i=1}^{N} p_t(i;\vh) \alpha_{t,i}.
	\end{align*}
	A direct computation yields that
	\begin{align*}
		\innerR{\nabla^2\overbar L_T(\vh)[\vz]}{\vz}
		&=
		\frac4{\tau^2T}\sum_{t=1}^T
		\left\{
		\sum_{i=1}^Np_t(i;\vh)\alpha_{t,i}^2
		-
		\left(\sum_{i=1}^Np_t(i;\vh)\alpha_{t,i}\right)^2
		\right\}\\
		&=\frac{2}{\tau^2 T} \sum_{t=1}^{T} \left\{ \sum_{i=1}^{N} \sum_{j=1}^{N} p_t(i;\vh) p_t(j;\vh) (\alpha_{t,i} - \alpha_{t,j})^2 \right\}\\
		&=\frac1{2\tau^2T}
		\sum_{t=1}^T\sum_{i=1}^N\sum_{j=1}^N
		p_t(i;\vh)p_t(j;\vh)
		\innerR{\mA_{t,i}-\mA_{t,j}}
		{\vz\vh^{\tranH}+\vh\vz^{\tranH}}^2.	
	\end{align*}
	where the second line is due to he elementary variance identity
	\begin{align*}
		\sum_i p_i\alpha_i^2-
		\left(\sum_i p_i\alpha_i\right)^2
		=
		\frac12\sum_{i=1}^N\sum_{j=1}^Np_ip_j(\alpha_i-\alpha_j)^2,
	\end{align*}
	and the lase line follows from the Hermitian identity
	\begin{align*}
		\innerR{\mA_{t,i}-\mA_{t,j}}
		{\vz\vh^{\tranH}+\vh\vz^{\tranH}}
		=
		2(\alpha_{t,i}-\alpha_{t,j}).
	\end{align*}

\subsection{Proof of Lemma~\ref{lem:phase-procrustes-alignment}}

	For $\phi\in[0,2\pi)$,
	\begin{align*}
		\|e^{-\mathrm i\phi}\vx-\vh\|_2^2
		=
		\|\vx\|_2^2+\|\vh\|_2^2
		-2\operatorname{Re}\!\left(
		e^{-\mathrm i\phi}\vh^{\tranH}\vx
		\right).
	\end{align*}
	If $\vh^{\tranH}\vx\neq0$, the stated choice of $\phi_{\vx}$ maximizes the
	last real part and gives
	$\vh^{\tranH}\vx^{\mathrm a}=|\vh^{\tranH}\vx|$.  If
	$\vh^{\tranH}\vx=0$, every $\phi\in[0,2\pi)$ is optimal and the same
	identity holds.  Consequently,
	\begin{align*}
		\vh^{\tranH}(\vx^{\mathrm a}-\vh)
		=
		|\vh^{\tranH}\vx|-\|\vh\|_2^2
		\in\R,
	\end{align*}
	which proves the horizontal-space claim.  The distance identity follows from
	the optimality of $\phi_{\vx}$.

\subsection{Proof of Lemma~\ref{lem:lifted-log-partition}}
	Fix $t$, a Hermitian matrix $\mM$, and a Hermitian direction $\mU$.  The chain
	rule gives
	\begin{align*}
		Df_t(\mM)[\mU]
		&=
		\frac1\tau
		\sum_{i=1}^Nq_t(i;\mM)
		\innerR{\mA_{t,i}}{\mU}\\
		&=
		\innerR{
			\frac1\tau
			\sum_{i=1}^Nq_t(i;\mM)\mA_{t,i}
		}{\mU}.
	\end{align*}
	The definition of the real gradient therefore proves the first identity.
	Next,
	\begin{align*}
		q_t(i;\mM)
		=
		\exp\left\{
		\frac1\tau\innerR{\mA_{t,i}}{\mM}-f_t(\mM)
		\right\},
	\end{align*}
	so
	\begin{align*}
		Dq_t(i;\mM)[\mU]
		&=
		\frac{q_t(i;\mM)}{\tau}
		\left[
		\innerR{\mA_{t,i}}{\mU}
		-
		\sum_{j=1}^Nq_t(j;\mM)
		\innerR{\mA_{t,j}}{\mU}
		\right]\\
		&=
		\frac{q_t(i;\mM)}{\tau}
		\sum_{j=1}^Nq_t(j;\mM)
		\innerR{\mA_{t,i}-\mA_{t,j}}{\mU}.
	\end{align*}
	Differentiating the gradient gives
	\begin{align*}
		\nabla^2f_t(\mM)[\mU]
		=
		\frac1{\tau^2}
		\sum_{i=1}^N\sum_{j=1}^N
		q_t(i;\mM)q_t(j;\mM)
		\innerR{\mA_{t,i}-\mA_{t,j}}{\mU}
		\mA_{t,i}.
	\end{align*}
	Interchanging the dummy indices $i$ and $j$ gives the equivalent expression
	with $-\mA_{t,j}$ in place of $\mA_{t,i}$.  Averaging the two expressions
	proves the symmetric Hessian formula.  Taking its real inner product with
	$\mU$ proves \eqref{eq:lifted-hessian}.

	\subsection{Proof of Lemma~\ref{lem:fisher-frame-consequences}}
		Set $\mH=\vh\vh^{\tranH}$, $\mX=\vx\vx^{\tranH}$, and
		$\bm\Delta=\mX-\mH$.  Since
		$p_t(i;\vh)=q_t(i;\mH)$, the population excess risk is the
		log-partition Bregman divergence
		\begin{align*}
			\overbar L_T(\vx)-\overbar L_T(\vh)
			&=
			\frac1T\sum_{t=1}^T
			\{f_t(\mX)-f_t(\mH)
			-\innerR{\nabla f_t(\mH)}{\bm\Delta}\}\\
			&=
			\frac1T\sum_{t=1}^T\int_0^1(1-s)
			\innerR{\nabla^2f_t(\mH+s\bm\Delta)[\bm\Delta]}
			{\bm\Delta}\,ds.
		\end{align*}
		For every $s\in[0,1]$,
		\begin{align*}
			\mH+s\bm\Delta
			=
			(1-s)\vh\vh^{\tranH}+s\vx\vx^{\tranH}
		\end{align*}
		is a convex combination of rank-one lifts generated by vectors with
		norms at most $R$.  Combining
		\eqref{eq:lifted-hessian} with the two inequalities in
		\eqref{eq:fisher-frame}, and using
		$\int_0^1(1-s)\,ds=1/2$, proves
		\eqref{eq:weighted-risk-sandwich}.

	\subsection{Proof of Lemma~\ref{lem:rank-one-quotient}}
		Let $\vx^{\mathrm a}$ be the phase-aligned representative from
		Lemma~\ref{lem:phase-procrustes-alignment}, and set
		$\vu=\vx^{\mathrm a}-\vh$ and $\vv=\vx^{\mathrm a}+\vh$.  Then
		$\|\vu\|_2=\distT(\vx,\vh)$ and
		\begin{align*}
			\vx\vx^{\tranH}-\vh\vh^{\tranH}
			=
			\frac12(\vu\vv^{\tranH}+\vv\vu^{\tranH}).
		\end{align*}
		Phase alignment makes $\vu^{\tranH}\vv=\|\vx\|_2^2-\|\vh\|_2^2$ real.
		Consequently,
		\begin{align*}
			\fronorm{\vx\vx^{\tranH}-\vh\vh^{\tranH}}^2
			&=
			\frac12\|\vu\|_2^2\|\vv\|_2^2
			+\frac12\operatorname{Re}\{(\vu^{\tranH}\vv)^2\}\\
			&\geq
			\frac12\|\vu\|_2^2\|\vv\|_2^2
			\geq
			\frac12\|\vh\|_2^2\distT^2(\vx,\vh),
		\end{align*}
		where $\|\vv\|_2^2=\|\vx\|_2^2+\|\vh\|_2^2
		+2|\vx^{\tranH}\vh|\geq\|\vh\|_2^2$.

	\subsection{Proof of Lemma~\ref{lem:horizontal-fisher-frame}}
		The symmetry of the real Hessian bilinear form restricted to
		$\mathcal H_{\vh}$ implies that $\mathcal I_{\vh}$ is self-adjoint.
		For $\vz\in\mathcal H_{\vh}$, put
		$\mathcal D_{\vh}(\vz)=\vz\vh^{\tranH}+\vh\vz^{\tranH}$.  The pairwise
		Hessian formula gives
		\begin{align*}
			\|\vz\|_{\mathcal I_{\vh}}^2
			=
			\frac1{2\tau^2T}\sum_{t=1}^T\sum_{i,j=1}^N
			p_t(i;\vh)p_t(j;\vh)
			\innerR{\mA_{t,i} - \mA_{t,j}}{\mathcal D_{\vh}(\vz)}^2.
		\end{align*}
		We obtain the frame bounds for this tangent direction directly from
		Assumption~\ref{assump:fisher-frame}, keeping
		$\mM=\vh\vh^{\tranH}$ fixed.  If $\|\vh\|_2<R$, set
		\begin{align*}
			\vy_\epsilon:=\vh,
			\qquad
			\vx_\epsilon:=\vh+\epsilon\vz.
		\end{align*}
		If $\|\vh\|_2=R$, choose
		$b>\max\{0,\operatorname{Re}(\vh^{\tranH}\vz)/R^2\}$ and set
		\begin{align*}
			\vy_\epsilon:=(1-b\epsilon)\vh,
			\qquad
			\vx_\epsilon:=\vy_\epsilon+\epsilon\vz.
		\end{align*}
		In either case, $\|\vx_\epsilon\|_2,\|\vy_\epsilon\|_2\leq R$ for all
		sufficiently small $\epsilon>0$.  Hence
		\begin{align*}
			\bm\Delta_\epsilon
			:=
			\vx_\epsilon\vx_\epsilon^{\tranH}
			-
			\vy_\epsilon\vy_\epsilon^{\tranH}
		\end{align*}
		is an admissible secant in Assumption~\ref{assump:fisher-frame}, and
		\begin{align*}
			\epsilon^{-1}\bm\Delta_\epsilon
			=
			\vz\vy_\epsilon^{\tranH}
			+\vy_\epsilon\vz^{\tranH}
			+\epsilon\vz\vz^{\tranH}
			\longrightarrow
			\mathcal D_{\vh}(\vz).
		\end{align*}
		Apply Assumption~\ref{assump:fisher-frame} to $\bm\Delta_\epsilon$ with
		$\mM=\vh\vh^{\tranH}$, divide by $\epsilon^2$, and let
		$\epsilon\downarrow0$.  Since
		$q_t(i;\vh\vh^{\tranH})=p_t(i;\vh)$, this yields
		\begin{align*}
			\kappa_F^-\fronorm{\mathcal D_{\vh}(\vz)}^2
			&\leq
			\frac1T\sum_{t=1}^T\sum_{i,j=1}^N
			p_t(i;\vh)p_t(j;\vh)
			\innerR{\mB_{t,ij}}{\mathcal D_{\vh}(\vz)}^2
			\leq
			\kappa_F^+\fronorm{\mathcal D_{\vh}(\vz)}^2.
		\end{align*}
		Moreover, horizontality makes
		$\vh^{\tranH}\vz$ real, so
		\begin{align*}
			2\|\vh\|_2^2\|\vz\|_2^2
			\leq
			\fronorm{\mathcal D_{\vh}(\vz)}^2
			=
			2\|\vh\|_2^2\|\vz\|_2^2
			+2(\vh^{\tranH}\vz)^2
			\leq
			4\|\vh\|_2^2\|\vz\|_2^2.
		\end{align*}
		The lower and upper frame inequalities now imply
		\eqref{eq:fisher-eigen-frame}.
		Since $\kappa_F^->0$ and $\|\vh\|_2>0$, the lower bound shows that
		$\mathcal I_{\vh}$ is positive definite on $\mathcal H_{\vh}$, so its
		quadratic form defines the stated norm.

	\subsection{Proof of Lemma~\ref{lem:score-concentration}}
	By \eqref{eq:sample-gradient}, each score is horizontal and centered:
\[
\nabla \ell_t(i;\bm h)\in \mathcal H_{\bm h},
\quad
\mathbb E\!\left\{
\nabla \ell_t(I_t;\bm h)
\right\}
=0.
\]
Moreover, for every $\bm z\in\mathcal H_{\bm h}$,
\[
\begin{aligned}
\frac{1}{T}\sum_{t=1}^T
\mathbb E\!\left[
\left\langle
\nabla \ell_t(I_t;\bm h),
\bm z
\right\rangle_{\mathbb R}^2
\right]
=
\frac{4}{\tau^2 T}
\sum_{t=1}^T
\operatorname{Var}_{I_t\sim p_t(\cdot;\bm h)}
\left(
\left\langle
\bm A_{t,I_t}\bm h,
\bm z
\right\rangle_{\mathbb R}
\right)
=
\left\langle
\nabla^2 \overline L_T(\bm h)[\bm z],
\bm z
\right\rangle_{\mathbb R}
=
\left\langle
\bm I_{\bm h}\bm z,
\bm z
\right\rangle_{\mathbb R},
\end{aligned}
\]
where the second equality follows from a intermediate result when proving \eqref{eq:population-hessian-pairwise}. Consequently, for every unit
vector $\bm u\in\mathcal H_{\bm h}$,
\[
\frac{1}{T}\sum_{t=1}^T
\mathbb E\!\left[
\left\langle
\bm u,
\bm I_{\bm h}^{-1/2}
\nabla \ell_t(I_t;\bm h)
\right\rangle_{\mathbb R}^2
\right]
=1.
\]
By Assumption~\ref{assump:bounded-design} and \eqref{eq:sample-gradient},
\[
\left\|
\bm I_{\bm h}^{-1/2}
\nabla \ell_t(i;\bm h)
\right\|_2
\le
\frac{2R}{\tau\sqrt{\lambda_-}}.
\]
Scalar Bernstein's inequality therefore gives, for every $x>0$,
\[
\mathbb P\left\{
\left|
\left\langle
\bm u,
\bm I_{\bm h}^{-1/2}
\nabla L_T(\bm h)
\right\rangle_{\mathbb R}
\right|
\ge
\sqrt{\frac{2x}{T}}
+
\frac{2Rx}{3\tau\sqrt{\lambda_-}T}
\right\}
\le
2e^{-x}.
\]

Applying this bound to a $1/2$-net of the unit sphere of the
$(2d-1)$-dimensional real space $\mathcal H_{\bm h}$, followed by a union
bound, yields
\[
\left\|
\bm I_{\bm h}^{-1/2}
\nabla L_T(\bm h)
\right\|_2
\le
2\left[
\sqrt{
\frac{
2\{2d\log 5+\log(2/\delta)\}
}{T}
}
+
\frac{
2R\{2d\log 5+\log(2/\delta)\}
}{
3\tau\sqrt{\lambda_-}T
}
\right]
\]
with probability at least $1-\delta$. Under the stated sample-size condition,
the second term is bounded by a constant multiple of the first. Squaring
completes the proof.

	\subsection{Proof of Lemma~\ref{lem:local-rsc}}
	It suffices to work directly on the aligned tube
$\mathcal N_\rho(\vh)$.  We divide the proof into four steps.
\begin{itemize}
	\item \textbf{Lipschitz continuity of the softmax weights.}
	For $\vu\in\R^N$, let
	\begin{align*}
		\sigma_i(\vu)
		:=
		\frac{e^{u_i}}{\sum_{j=1}^Ne^{u_j}},
		\quad i\in[N].
	\end{align*}
	The Jacobian entries satisfy
	\begin{align*}
		\frac{\partial\sigma_i(\vu)}{\partial u_j}
		=
		\sigma_i(\vu)\bigl(\delta_{ij}-\sigma_j(\vu)\bigr).
	\end{align*}
	Consequently, for every $\vb\in\R^N$,
	\begin{align*}
		\nabla\sigma_i(\vu)^{\mathsf T}\vb
		&=
		\sigma_i(\vu)
		\left(
		b_i-\sum_{j=1}^N\sigma_j(\vu)b_j
		\right),\\
		\sum_{i=1}^N
		\left|\nabla\sigma_i(\vu)^{\mathsf T}\vb\right|
		&\leq
		\sum_{i=1}^N\sigma_i(\vu)|b_i|
		+
		\sum_{i=1}^N\sigma_i(\vu)
		\sum_{j=1}^N\sigma_j(\vu)|b_j|\leq
		2\|\vb\|_\infty.
	\end{align*}
	Applying the fundamental theorem of calculus along the segment from
	$\vv$ to $\vu$ gives
	\begin{align*}
		\sum_{i=1}^N|\sigma_i(\vu)-\sigma_i(\vv)|
		&\leq
		\int_0^1
		\sum_{i=1}^N
		\left|
		\nabla\sigma_i\bigl(\vv+s(\vu-\vv)\bigr)^{\mathsf T}
		(\vu-\vv)
		\right|\,ds\leq
		2\|\vu-\vv\|_\infty.
	\end{align*}
	We apply this inequality to the logit vectors associated with
	$\vx,\vx'\in\C^d$ satisfying $\|\vx\|_2,\|\vx'\|_2\leq R$.
	Assumption~\ref{assump:bounded-design} gives
	$|\va_{t,i}^{\tranH}\vx|,|\va_{t,i}^{\tranH}\vx'|\leq R$, and hence
	\begin{align*}
		\left|
		|\va_{t,i}^{\tranH}\vx|^2
		-|\va_{t,i}^{\tranH}\vx'|^2
		\right|
		&\leq
		\left(
		|\va_{t,i}^{\tranH}\vx|
		+|\va_{t,i}^{\tranH}\vx'|
		\right)
		\left|
		|\va_{t,i}^{\tranH}\vx|
		-|\va_{t,i}^{\tranH}\vx'|
		\right|\\
		&\leq
		2R|\va_{t,i}^{\tranH}(\vx-\vx')|\leq
		2R\|\vx-\vx'\|_2.
	\end{align*}
	Therefore
	\begin{align*}
		\sum_{i=1}^N|p_t(i;\vx)-p_t(i;\vx')|
		&\leq
		\frac2\tau
		\max_{i\in[N]}
		\left|
		|\va_{t,i}^{\tranH}\vx|^2
		-|\va_{t,i}^{\tranH}\vx'|^2
		\right|\leq
		\frac{4R}{\tau}\|\vx-\vx'\|_2.
	\end{align*}
	
	\item \textbf{Lipschitz continuity of the population Hessian.}
	We next bound each of the three point-dependent terms in
	\eqref{eq:population-hessian}.  
	\begin{itemize}
		\item By the definition of $\mC_t$ and
		$\opnorm{\mA_{t,i}}\leq1$,
		\begin{align*}
			\opnorm{\mC_t(\vx)-\mC_t(\vx')}
			&\leq
			\sum_{i=1}^N
			|p_t(i;\vx)-p_t(i;\vx')|
			\opnorm{\mA_{t,i}}\leq
			\frac{4R}{\tau}\|\vx-\vx'\|_2.
		\end{align*}
		
		\item For every $\vz\in\C^d$ with $\|\vz\|_2=1$, a simple computation yields
		\begin{align*}
			\mS_t(\vx)[\vz]-\mS_t(\vx')[\vz]
			&=
			\sum_{i=1}^N
			\bigl(p_t(i;\vx)-p_t(i;\vx')\bigr)
			\innerR{\mA_{t,i}\vx}{\vz}\mA_{t,i}\vx\\
			&\quad+
			\sum_{i=1}^Np_t(i;\vx')
			\innerR{\mA_{t,i}\vx}{\vz}
			\mA_{t,i}(\vx-\vx')\\
			&\quad+
			\sum_{i=1}^Np_t(i;\vx')
			\innerR{\mA_{t,i}(\vx-\vx')}{\vz}
			\mA_{t,i}\vx'.
		\end{align*}
		Since $\|\mA_{t,i}\vx\|_2,\|\mA_{t,i}\vx'\|_2\leq R$, one has
		\begin{align*}
			\|\mS_t(\vx)[\vz]-\mS_t(\vx')[\vz]\|_2 &\leq			R^2\sum_{i=1}^N|p_t(i;\vx)-p_t(i;\vx')|+ 2R\|\vx-\vx'\|_2\\
			&\leq
			\left(\frac{4R^3}{\tau}+2R\right)
			\|\vx-\vx'\|_2.
		\end{align*}
		Taking the supremum over all unit directions $\vz$ gives
		\begin{align*}
			\sup_{\|\vz\|_2=1}
			\|\mS_t(\vx)[\vz]-\mS_t(\vx')[\vz]\|_2
			\leq
			\left(\frac{4R^3}{\tau}+2R\right)
			\|\vx-\vx'\|_2.
		\end{align*}
		
		\item The definition $\vy_t(\vx)=\mC_t(\vx)\vx$ gives
		\begin{align*}
			\vy_t(\vx)-\vy_t(\vx')
			&= \mC_t(\vx) \vx - \mC_t(\vx') \vx'=\sum_{i=1}^{N} p_t(i; \vx) \mA_{t,i} \vx - \sum_{i=1}^{N} p_t(i; \vx') \mA_{t,i} \vx'\\
			&=	\sum_{i=1}^N\bigl(p_t(i;\vx)-p_t(i;\vx')\bigr)			\mA_{t,i}\vx+\sum_{i=1}^Np_t(i;\vx')\mA_{t,i}(\vx-\vx'),
		\end{align*}
		and hence
		\begin{align*}
			\|\vy_t(\vx)-\vy_t(\vx')\|_2
			&\leq
			R\sum_{i=1}^N|p_t(i;\vx)-p_t(i;\vx')|
			+\|\vx-\vx'\|_2\leq
			\left(1+\frac{4R^2}{\tau}\right)
			\|\vx-\vx'\|_2.
		\end{align*}
		Moreover, $\|\vy_t(\vx)\|_2,\|\vy_t(\vx')\|_2\leq R$.
		For every unit vector $\vz$, one has
		\begin{align*}
			\innerR{\vy_t(\vx)}{\vz}\vy_t(\vx)
			-
			\innerR{\vy_t(\vx')}{\vz}\vy_t(\vx')&=
			\innerR{\vy_t(\vx)}{\vz}
			\bigl(\vy_t(\vx)-\vy_t(\vx')\bigr)
			+
			\innerR{\vy_t(\vx)-\vy_t(\vx')}{\vz}
			\vy_t(\vx').
		\end{align*}
		It follows that
		\begin{align*}
			\sup_{\|\vz\|_2=1}
			\left\|
			\innerR{\vy_t(\vx)}{\vz}\vy_t(\vx)
			-
			\innerR{\vy_t(\vx')}{\vz}\vy_t(\vx')
			\right\|_2&\leq
			\left(\|\vy_t(\vx)\|_2+\|\vy_t(\vx')\|_2\right)
			\|\vy_t(\vx)-\vy_t(\vx')\|_2\\
			&\leq
			\left(2R+\frac{8R^3}{\tau}\right)
			\|\vx-\vx'\|_2.
		\end{align*}
	\end{itemize}
	We now combine these bounds directly at the level of quadratic forms:
	\begin{align*}
		&\sup_{\|\vz\|_2=1}
		\left|
		\innerR{
			\bigl(
			\nabla^2\overbar L_T(\vx)-\nabla^2\overbar L_T(\vx')
			\bigr)[\vz]
		}{\vz}
		\right|\\
		&=
		\sup_{\|\vz\|_2=1}
		\left|
		\frac2{\tau T}\sum_{t=1}^T
		\innerR{
			\bigl(\mC_t(\vx)-\mC_t(\vx')\bigr)\vz
		}{\vz}
		\right.\left.
		+
		\frac4{\tau^2T}\sum_{t=1}^T
		\innerR{
			\mS_t(\vx)[\vz]-\mS_t(\vx')[\vz]
		}{\vz}
		\right.\\
		&\quad\left.
		-
		\frac4{\tau^2T}\sum_{t=1}^T
		\innerR{
			\innerR{\vy_t(\vx)}{\vz}\vy_t(\vx)
			-
			\innerR{\vy_t(\vx')}{\vz}\vy_t(\vx')
		}{\vz}
		\right|\\
		&\leq 
		\frac2{\tau T}\sum_{t=1}^T
		\opnorm{\mC_t(\vx)-\mC_t(\vx')} +
		\frac4{\tau^2T}\sum_{t=1}^T
		\sup_{\|\vw\|_2=1}
		\|\mS_t(\vx)[\vw]-\mS_t(\vx')[\vw]\|_2\\
		&\quad+
		\frac4{\tau^2T}\sum_{t=1}^T
		\sup_{\|\vw\|_2=1}
		\left\|
		\innerR{\vy_t(\vx)}{\vw}\vy_t(\vx)
		-
		\innerR{\vy_t(\vx')}{\vw}\vy_t(\vx')
		\right\|_2\\
		&\leq
		\left[
		\frac2\tau\frac{4R}{\tau}
		+
		\frac4{\tau^2}\left(\frac{4R^3}{\tau}+2R\right)
		+
		\frac4{\tau^2}\left(2R+\frac{8R^3}{\tau}\right)
		\right]
		\|\vx-\vx'\|_2\\
		&=
		\left(
		\frac{48R^3}{\tau^3}
		+
		\frac{24R}{\tau^2}
		\right)
		\|\vx-\vx'\|_2\\
		&=
		L_H\|\vx-\vx'\|_2. \numberthis \label{eq:population-hessian-lipschitz}
	\end{align*}

	\item \textbf{Population curvature on the aligned tube.}
	Let $\vx\in\mathcal N_\rho(\vh)$ and $\vz\in\mathcal H_{\vh}$.
	By the definition of $\mathcal I_{\vh}$ in \eqref{eq:horizontal-fisher},
	\eqref{eq:population-hessian-lipschitz}, and
	$\|\vx-\vh\|_2\leq\rho\leq\lambda_-/(8L_H)$,
	\begin{align*}
		\innerR{\nabla^2\overbar L_T(\vx)[\vz]}{\vz}
		&=
		\|\vz\|_{\mathcal I_{\vh}}^2
		+
		\innerR{
			\bigl(\nabla^2\overbar L_T(\vx)-\nabla^2\overbar L_T(\vh)\bigr)[\vz]
		}{\vz}\\
		&\geq
		\|\vz\|_{\mathcal I_{\vh}}^2
		-
		\sup_{\|\vw\|_2=1}
		\left|
		\innerR{
			\bigl(
			\nabla^2\overbar L_T(\vx)-\nabla^2\overbar L_T(\vh)
			\bigr)[\vw]
		}{\vw}
		\right|
		\|\vz\|_2^2\\
		&\geq
		\|\vz\|_{\mathcal I_{\vh}}^2-L_H\rho\|\vz\|_2^2
		\geq
		\frac78\|\vz\|_{\mathcal I_{\vh}}^2.
		\numberthis\label{eq:nearby-population-rsc1}
	\end{align*}

	\item \textbf{Empirical curvature on the aligned tube.}
	Subtracting \eqref{eq:population-hessian} from
	\eqref{eq:sample-hessian} gives
	\begin{align*}
		\nabla^2L_T(\vx)-\nabla^2\overbar L_T(\vx)&=
		-\frac2{\tau T}\sum_{t=1}^T
		\bigl(\mA_{t,I_t}-\mC_t(\vh)\bigr):=
		-\frac2{\tau T}\sum_{t=1}^T\mZ_t,
	\end{align*}
	where
	\begin{align*}
		\mZ_t
		=
		\mA_{t,I_t}-\mC_t(\vh)
		=
		\mA_{t,I_t}-\E{\mA_{t,I_t}}.
	\end{align*}
	The matrices $\mZ_1,\ldots,\mZ_T$ are independent, mean-zero, and
	Hermitian.  Since
	\begin{align*}
		\bm0\preceq\mA_{t,I_t}\preceq\mI_d,
		\quad
		\bm0\preceq\mC_t(\vh)\preceq\mI_d,
	\end{align*}
	one has $-\mI_d\preceq\mZ_t\preceq\mI_d$ and therefore
	\(
	\opnorm{\mZ_t}\leq 1.
	\)
	Moreover,
	\begin{align*}
		\E{\mZ_t^2}=
		\E{\mA_{t,I_t}^2}
		-\mC_t(\vh)^2\preceq
		\E{\mA_{t,I_t}^2}\preceq
		\E{\mA_{t,I_t}}
		=
		\mC_t(\vh)
		\preceq
		\mI_d,
	\end{align*}
	where $\mA_{t,I_t}^2\preceq\mA_{t,I_t}$ follows from
	$\bm0\preceq\mA_{t,I_t}\preceq\mI_d$.  Hence
	\begin{align*}
		\opnorm{\sum_{t=1}^T\E{\mZ_t^2}}
		\leq T.
	\end{align*}
	The complex Hermitian form of matrix Bernstein's inequality therefore
	implies that, with probability at least $1-\delta$,
	\begin{align*}
		\opnorm{\sum_{t=1}^T\mZ_t}
		\leq
		\sqrt{2T\log(2d/\delta)}
		+
		\frac23\log(2d/\delta).
	\end{align*}
	On this event, the Hessian fluctuation is independent of $\vx$.  Moreover,
	for every $\vz\in\C^d$ with $\|\vz\|_2=1$,
	\begin{align*}
		\left|
		\innerR{
			\left(\sum_{t=1}^T\mZ_t\right)\vz
		}{\vz}
		\right|
		&=
		\left|
		\operatorname{Re}\left\{
		\vz^{\tranH}\left(\sum_{t=1}^T\mZ_t\right)\vz
		\right\}
		\right|\leq
		\left|
		\vz^{\tranH}\left(\sum_{t=1}^T\mZ_t\right)\vz
		\right|\leq
		\opnorm{\sum_{t=1}^T\mZ_t}.
	\end{align*}
	Consequently,
	\begin{align*}
		&\sup_{\vx\in\mathcal N_\rho(\vh)}
		\sup_{\substack{\vz\in\mathcal H_{\vh}\\\|\vz\|_2=1}}
		\left|
		\innerR{
			\bigl(\nabla^2L_T(\vx)-\nabla^2\overbar L_T(\vx)\bigr)[\vz]
		}{\vz}
		\right| \leq
		\frac{2}{\tau T}
		\opnorm{\sum_{t=1}^T\mZ_t}\\
	&\leq
	\frac{2\sqrt2}{\tau}
	\sqrt{\frac{\log(2d/\delta)}{T}}
	+
	\frac4{3\tau}
	\frac{\log(2d/\delta)}{T}\leq \frac{\lambda_-}{8},
	\end{align*}
	where the last line follows from the sample-size assumption for a
	sufficiently large absolute constant $C_{\mathrm{rsc}}$.
	Combining this estimate with \eqref{eq:nearby-population-rsc1}, for every
	$\vx\in\mathcal N_\rho(\vh)$ and $\vz\in\mathcal H_{\vh}$,
	\begin{align*}
		\innerR{\nabla^2L_T(\vx)[\vz]}{\vz}
		&\geq
		\innerR{\nabla^2\overbar L_T(\vx)[\vz]}{\vz}
		-
		\sup_{\substack{\vw\in\mathcal H_{\vh}\\\|\vw\|_2=1}}
		\left|
		\innerR{
			\bigl(\nabla^2L_T(\vx)-\nabla^2\overbar L_T(\vx)\bigr)[\vw]
		}{\vw}
		\right|
		\|\vz\|_2^2\\
		&\geq
		\frac78\|\vz\|_{\mathcal I_{\vh}}^2
		-
		\frac{\lambda_-}{8}\|\vz\|_2^2
		\geq
		\frac34\|\vz\|_{\mathcal I_{\vh}}^2.
	\end{align*}

\end{itemize}

\section{Conclusion and Future Work}
\label{sec conclusion}
% 多流的，实验work良好，分析上应该类似
% 未来工作, 直接分析quantize的performance
This work develops a non-asymptotic theory for the constrained MLE under 
the phaseless Gumbel--Softmax index model, a smooth probabilistic surrogate 
for the deterministic hard-PMI selection rule. The analysis connects global 
excess-risk control with phase-aligned localization and local curvature, 
providing a unified path from likelihood concentration to sharp channel-recovery 
guarantees. Together with the local minimax lower bound, these results identify 
the main statistical scaling of the problem up to design-dependent factors. 
Experiments on QuaDRiGa-generated FDD channels further show that the resulting 
likelihood-based estimator remains effective when applied directly to 
deterministic hard-PMI observations. Overall, the results demonstrate that the Gumbel--Softmax formulation 
is analytically tractable while retaining practical value for PMI-only channel 
recovery.

There are  two  interesting directions for future work. Firstly, it is likely to extend the present non-asymptotic theory to matrix-valued channels with multi-stream feedback. While the likelihood-based framework applies directly in this setting, the corresponding theory requires replacing the global-phase ambiguity by a right-unitary invariance and developing the associated quotient geometry and local identifiability analysis. Secondly, deterministic hard-PMI feedback merits a direct theoretical 
analysis. Establishing finite-sample upper and lower error bounds would 
characterize the fundamental recovery limit under hard-PMI observations and 
guide the development of algorithms that attain this limit.
%%%%%%%%%%%%%%%%%%%

\bibliographystyle{plain}
\bibliography{refs}
\end{document}